\documentclass[%
 reprint,
 amsmath,amssymb,
 aps,
]{revtex4-2}

\usepackage{graphicx}
\usepackage{dcolumn}
\usepackage{bm}
\usepackage{float}

\begin{document}


\title{On the stability of traffic breakup patterns in urban networks}

\author{Marco Cogoni}
\email{marco.cogoni@crs4.it}
\author{Giovanni Busonera}%
\affiliation{CRS4: Centro di Ricerche, Sviluppo e Studi Superiori in Sardegna, Ex-distilleria, Via Ampere, 2 - 09134 Cagliari (CA), ITALY}%

\date{\today}

\begin{abstract}
We investigate the behavior of extended urban traffic networks within the framework of percolation theory by using real and synthetic traffic data. 
Our main focus shifts from the statistical properties of the cluster size distribution studied recently, to the spatial analysis of the clusters at criticality and to the definition of a similarity measure between whole urban configurations. We discover that the breakup patterns of the complete network, formed by the connected functional road clusters at criticality, show remarkable stability from one hour to the next, and predictability for different days at the same time. We prove this by showing how the average spatial distributions of the highest-rank clusters evolve over time, and by building a taxonomy of traffic states via dimensionality-reduction of the distance matrix, obtained via a clustering similarity score. Finally, we show that a simple random percolation model can approximate the breakup patterns of heavy real traffic when long-ranged spatial correlations are imposed. 
\end{abstract}

\maketitle

Traffic has been extensively studied in recent years with a focus on the free-flow to congestion transition. Several approaches have been employed ranging from analytical models to coarse-grained traffic descriptions~\cite{helbing_traffic_2001, kerner1998experimental, helbing1998coherent, chowdhury2000statistical}. Recently, a new perspective based on percolation theory~\cite{li_percolation_2015,zeng_switch_2019, zeng2020multiple}, has been proposed to study traffic flows in large cities by leveraging the availability of extensive datasets with road-level speed data with sufficient time resolution~\cite{UBER_Movement}. Within this approach, single-vehicle dynamics is absent and the focus is about roads being able to guarantee transportation efficiency above some threshold: road segments (bonds or edges in graph terminology) failing this criterion are removed from the graph representing the urban network. The theoretical setup is that of isotropic random bond percolation over irregular planar networks~\cite{de_Noronha_2018, callaway2000network}.

These studies proved that, beyond the well known passage from free-flow to the congested state happening at road level, a global percolation transition exists: by increasing the threshold speed, the network is progressively disrupted from a single giant connected component to a set of smaller road clusters. While each cluster is able to sustain internal traffic above the threshold speed, only slower connections are available when moving among different clusters~\cite{li_percolation_2015}. A critical speed is known to exist for which the size distribution of clusters follows a power law $n_s\sim s^{-\tau}$, where $s$ is the cluster size (measured in number of road segments), $n_s$ is the ratio between the number of $s$-sized clusters and the total number of clusters, and $\tau$ is the critical percolation exponent~\cite{li_percolation_2015}. The $\tau$ (Fisher) exponent is known to change under different traffic regimes by a factor determined by the specific city topology~\cite{zeng_switch_2019}.

While characterizing traffic by a single scalar with the critical exponent $\tau$ is useful to discover changes of universality class and to measure the influence of topologic features on them~\cite{zeng_switch_2019}, here we shift our focus from the statistical properties of the cluster size distribution~\cite{stauffer2018introduction} to the spatial analysis of the clusters at criticality and to the definition of a similarity measure between whole urban configurations, especially in order to estimate their predictability during equivalent traffic scenarios on different days. The same measure can be used to estimate the stability of the clustering configurations over time, during the same day.

Being able to objectively quantify why and where two traffic breakup patterns differ, by means of clustering similarity and geographical representations of the connectivity breakup patterns, leads to a novel powerful approach in the study of urban network dynamics, especially for the understanding of how and when congestion appears and dissolves with no need for details about the underlying vehicle dynamics. The clustering similarity information allows a direct and real-time classification of whole-city states, enabling the early detection of traffic anomalies.
To gain a deeper understanding of the underlying processes involved in the emergence of breakup patterns, we present a simple random percolation model to generate synthetic whole-city traffic configurations possessing properties compatible with those observed in the real-world dataset. One of the key aspects of our random percolation model is the possibility to generate spatial correlations with a power law decay on a graph. It turns out that only a specific range of spectral decay exponents leads to configurations producing similar breakup patterns to those seen in real traffic during certain time slots.

The generality of this approach could prove useful for the study of entirely different critical phenomena appearing in disparate settings such as energy grids, global air and maritime shipping networks, and social interactions: all lending themselves easily to percolation analysis.


{\it Methods ---} We describe the transportation networks by means of directed weighted graphs. Each graph possesses $N$ nodes and $M$ edges. Each edge encodes the real length of the corresponding road segment, which is, in general, larger than the euclidean distance between neighboring intersections. Edge weights are used only to compute the Laplacian eigenvalues, necessary to create spatially-correlated noise on the graph.

Percolation on our graph is performed by assigning a value $q_i\in(0,1)$ to the $i$-th edge for $i\le M$, that may be a real (normalized) speed or a random number, and then erasing all edges for which $q_i<q^*$, where $q^*$ is the chosen percolation threshold. Erased edges are called \emph{dysfunctional}, as opposed to the \emph{functional} ones surviving in the graph. There is a monotonic dependence between $q^*$ and the number of functional edges $M^f$. The ratio of these edges and $M$ is normally called $p=M^f/M$ in percolation theory. The threshold $q^*$ coincides with the probability of edge destruction $1-p$ when $q_i$ are distributed as uniform random noise.
Only one value of $p$ is associated to the critical percolation transition for which the size distribution of clusters follows a power law and it is called the critical probability $p_c$. To find $p_c$, we perform multiple percolations for a uniformly distributed set of $q^*\in(0,1)$ (with $0.01$ spacing), to find the critical value $q^*_c$ corresponding to the maximum size of the second largest cluster~\cite{li_percolation_2015, bastas2014method}.

In this work we use the transportation networks of the metropolitan areas of New York City and London \footnote{$1600$ km$^2$, only roads open to private cars} from OpenStreetMap (OSM)~\cite{OpenStreetMap}. Each street network was directly extracted from the OSM dataset with the open library OSMnX~\cite{OSMNX} and NetworkX was used to analyze graphs~\cite{networkx}.
Real velocity data (matched to the same OSM graph) was obtained from UBER~\cite{UBER_Movement} and it contains, with a one-hour granularity, the velocity recorded by GPS sensors installed on their taxis (average of several taxis on the same edge). During each hour there is a varying fraction of edges not containing any speed information that needs to be locally interpolated (see Supplemental Materials for details on the dataset and about preprocessing).

One of the goals of this work is to verify whether by inducing spatial correlations on random speed values $q_i$ leads to more realistic network configurations with respect to Bernoulli percolation~\cite{helbing1998coherent,helbing_traffic_2001,Wang_2015}. In order to achieve correlation on weighted graphs, we compute the approximate spectrum of the weighted graph Laplacian for the entire urban network \cite{defferrard_pygsp_2017}, then apply a Graph Fourier Transform to the uncorrelated noise and filter it~\cite{prakash_structural_1992} by multiplying the spectrum with a power law $f(k) \sim k^{{(d-\lambda)}/2}$: a rapid decay (i.e., $\lambda\rightarrow 0$, filters out the highest vibrational modes of the graph) leads to the longest spatial correlations \cite{makse_method_1996}, while $\lambda\rightarrow 2$ leaves the noise uncorrelated. Since the urban network is planar, $d=2$ is imposed.

We compute the size distribution of the functional clusters of the graph at criticality in the same way for random and real data: after pruning the dysfunctional edges, we extract the strongly connected components of the graph. Within these clusters traffic can move at speeds greater than the threshold $q^*$~\footnote{By computing the critical percolation transition considering the weakly connected components instead, we obtained very similar results, except for the critical speed threshold, as expected from existing results~\cite{de_Noronha_2018}.}.
To compute the critical exponent $\tau$, we perform a linear fit on a log-log plot of the binned distribution of the cluster sizes ($12$ bins for the domain $(0,4)$) for each percolation instance (obtained for each hour of each day) for real data, whereas $10^3$ instances obtained from random percolation were pooled together to decrease statistical error.

In order to demonstrate the average spatial stability of the largest clusters throughout the urban network when observed at equal hours for several months, we introduce the computation of the spatial distribution of the occurrences of each cluster (i.e., the largest, the second and the union of the smaller clusters up to the $1000$-th) over all graph edges. The final map is obtained by considering hundreds (for real data) or thousands (for synthetic velocities) of equivalent percolation instances and by binning the occurrences associated to nearby edges.

We use the Fowlkes-Mallows (FM) score~\cite{fowlkes_mallows1983} to quantify the similarity between two different clusterings $C$ and $C'$:
$${\rm FM}(C,C') = {n_{11}\over{\sqrt{(n_{11}+n_{10})(n_{11}+n_{01})}}}$$
where $n_{11}$ is the number of edges that belong to the same cluster both in $C$ and $C'$, $n_{10}$ is the number of edges that belong to the same cluster in $C$ but do not in  $C'$, and finally, $n_{01}$ is the number of edges that do not belong to the same cluster in $C$ but do in  $C'$.
Several measures have been proposed to quantify the similarity among different clustering outcomes\cite{meilua2007comparing}, here we decided to adopt the FM score because random independent clusterings produce ${\rm FM}\sim0$, whereas ${\rm FM}=1$ for perfect matchings, and for its symmetry~\footnote{We compute the FM score by considering the largest five clusters for each instance.}. Results obtained by means of other similarity measures (i.e., largest-cluster overlap, adjusted Rand score and Hamming distance) are qualitatively compatible with FM, albeit less than optimal for the task.

After computing the similarity matrix, we apply the t-SNE dimensionality reduction algorithm~\cite{van2008visualizing} to the similarity information (obtained at criticality) to get a simple classification of unknown traffic states. The alternative approach of directly comparing raw velocity data (instead of spatial breakup patterns), by computing the euclidean distance between two traffic states, may not detect crucial differences. This may be due, for instance, to a very small number of important roads possessing very different congestion levels: in which case the two breakup patterns would turn out completely different. Since the graphs representing large urban areas contain about $\sim10^5$ edges, traffic connectivity may be severely impaired by disrupting a comparatively small number of edges possessing a high betweenness centrality~\cite{kirkley2018betweenness}.

{\it Results ---} The Fisher critical exponent $\tau$, which characterizes the size distribution of functional clusters, has been recently used as an indicator to distinguish among the possible global states of urban traffic~\cite{zeng_switch_2019}. Its value is known to depend on traffic intensity, at least for specific topologies comprising high-speed urban highways (e.g., Beijing). For real data, we observe a very small change in the value of $\tau$ for both London and New York City ($\approx1.8$ for rush hours and $\approx1.9$ for off-peak hours) as shown in Table~\ref{tab:tau-exponent}. On the other hand, substantial differences are observed for $q_c$ whether considering rush hours or not for both cities, as seen in Table~\ref{tab:qc}. Theoretical results~\cite{de_Noronha_2018} for randomly directed (isotropic) percolation on planar graphs show a higher value of $\tau \approx 2.1$. Random percolation with long-ranged spatial correlations ($\lambda\rightarrow0$) show a lower value of $\tau$, similar to what we see for rush-hours, while Bernoulli percolation ($\lambda\rightarrow2$) leads to $\tau$ values between $2.02$ and $2.05$, similar to the theoretical result~\cite{verbavatz2021oneway}, well outside the range observed for real traffic. The dependence of $\tau$ on $\lambda$ is a smooth and monotonic function.

\begin{table}[h!]
\begin{center}
\begin{tabular}{ cccccccc } 
  & $8am$ & $10am$ & $5pm$ & $8pm$ & RND & LNG & TH\\
 \hline
 London & $1.82(82)$ & $1.86(84)$ & $1.82(85)$ & $1.85(85)$ & $2.05$ & $1.85$ & $2.1$\\ 
 NYC    & $1.80(89)$ & $1.87(91)$ & $1.80(82)$ & $1.89(87)$ & $2.02$ & $1.73$ & $2.1$\\ 
\hline
\end{tabular}
\caption{Average over three months for $\tau$ for London and NYC during weekdays (within parentheses decimals for weekends). UBER data at four different hours, random percolation with long (LNG) and without (RND) spatial correlation, and the theoretical prediction for planar graphs with randomly (isotropic) directed bonds~\cite{de_Noronha_2018}.}
\label{tab:tau-exponent}

\begin{tabular}{ cccccccc } 
  & $8am$ & $10am$ & $5pm$ & $8pm$\\
 \hline
 London & $0.41(60)$ & $0.46(51)$ & $0.40(45)$ & $0.52(52)$\\ 
 NYC    & $0.46(64)$ & $0.56(60)$ & $0.44(50)$ & $0.58(56)$\\ 
 \hline
\end{tabular}
\caption{Average over three months for $q_c$ for London and NYC during weekdays (within parentheses decimals for weekends). UBER data at four different hours.}
\label{tab:qc}
\end{center}
\end{table}

In Table~\ref{tab:tau-exponent}, the observed $\tau$ changes are small, but consistent, for both cities when comparing peak and off-peak hours. These results seem to confirm previous findings~\cite{zeng_switch_2019} affirming that the absence of an extended network of urban highways, acting as traffic shortcuts among distant locations, leads to small, if any, variations of $\tau$. We observe that $\tau\approx2$ at all times, corresponding to a standard Zipf's Law or to the maximal value of the diversity index~\cite{mazzarisi2021maximal}, except for heavy congestion (during rush hours or random percolation with long-ranged spatial correlation), when we observe a reduction in the diversity of cluster sizes. This diversity index depends on the specific city topology~\cite{verbavatz2021oneway}.


\begin{figure*}[ht!]
\centering
    \includegraphics[width=0.24\textwidth]{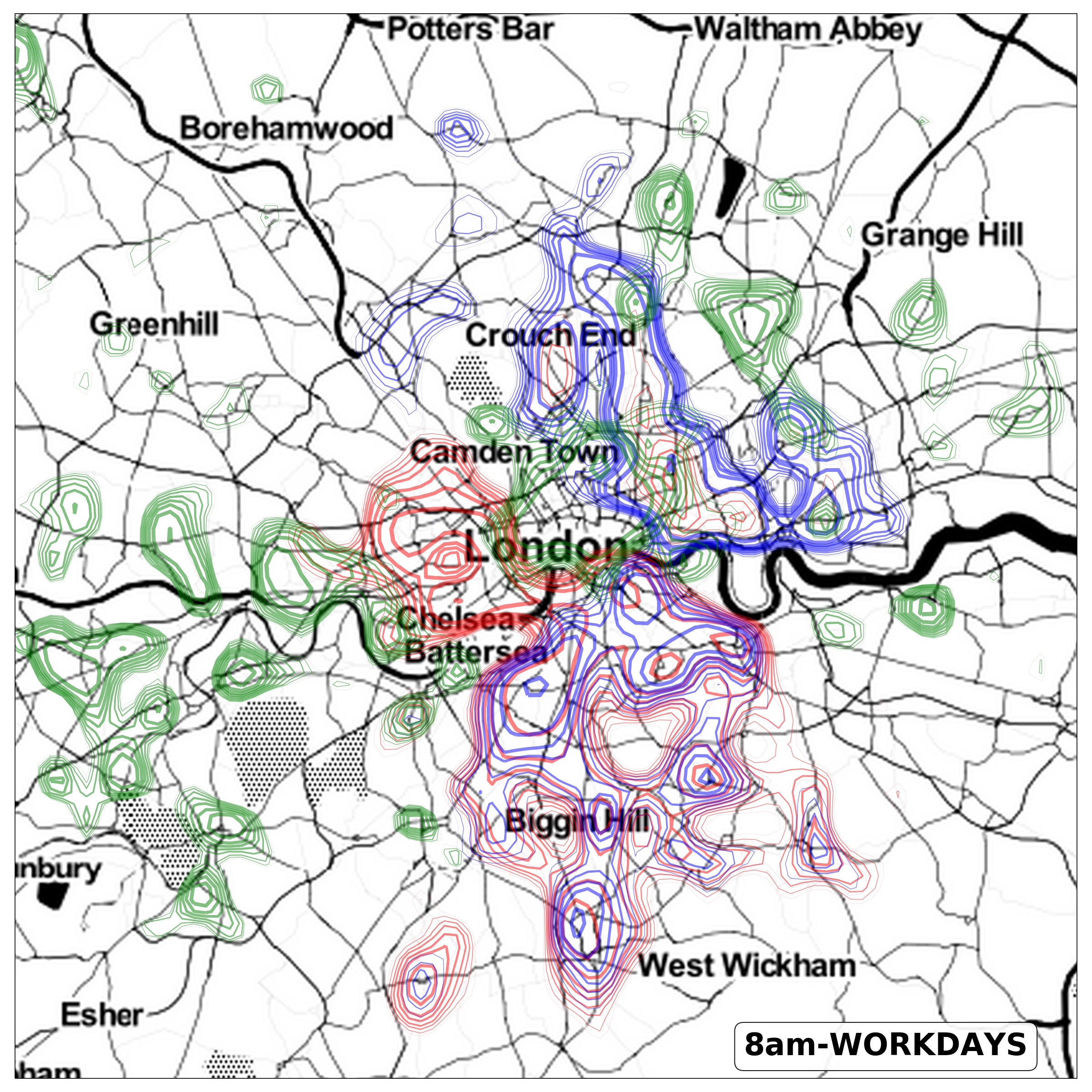}
    \includegraphics[width=0.24\textwidth]{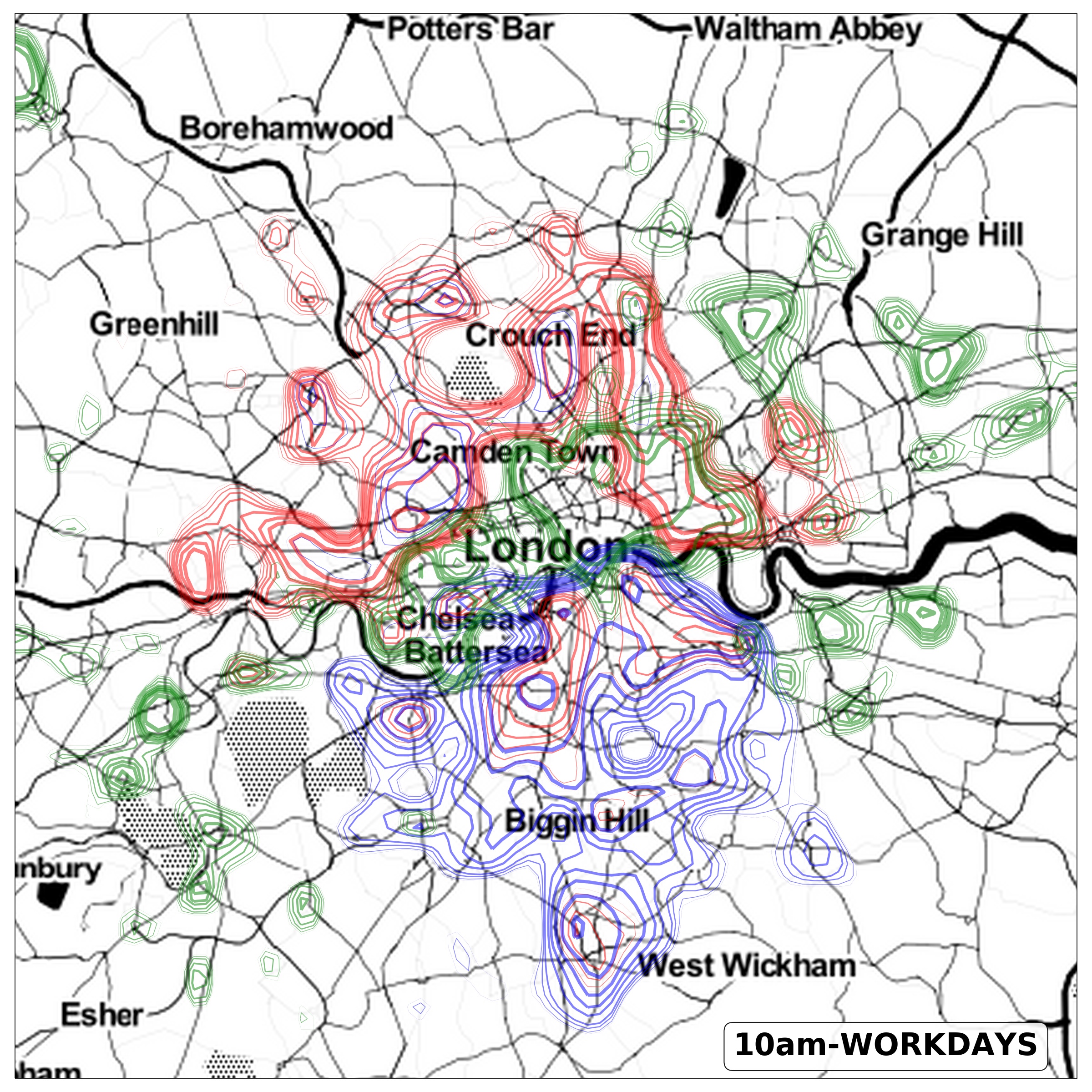}
    \includegraphics[width=0.24\textwidth]{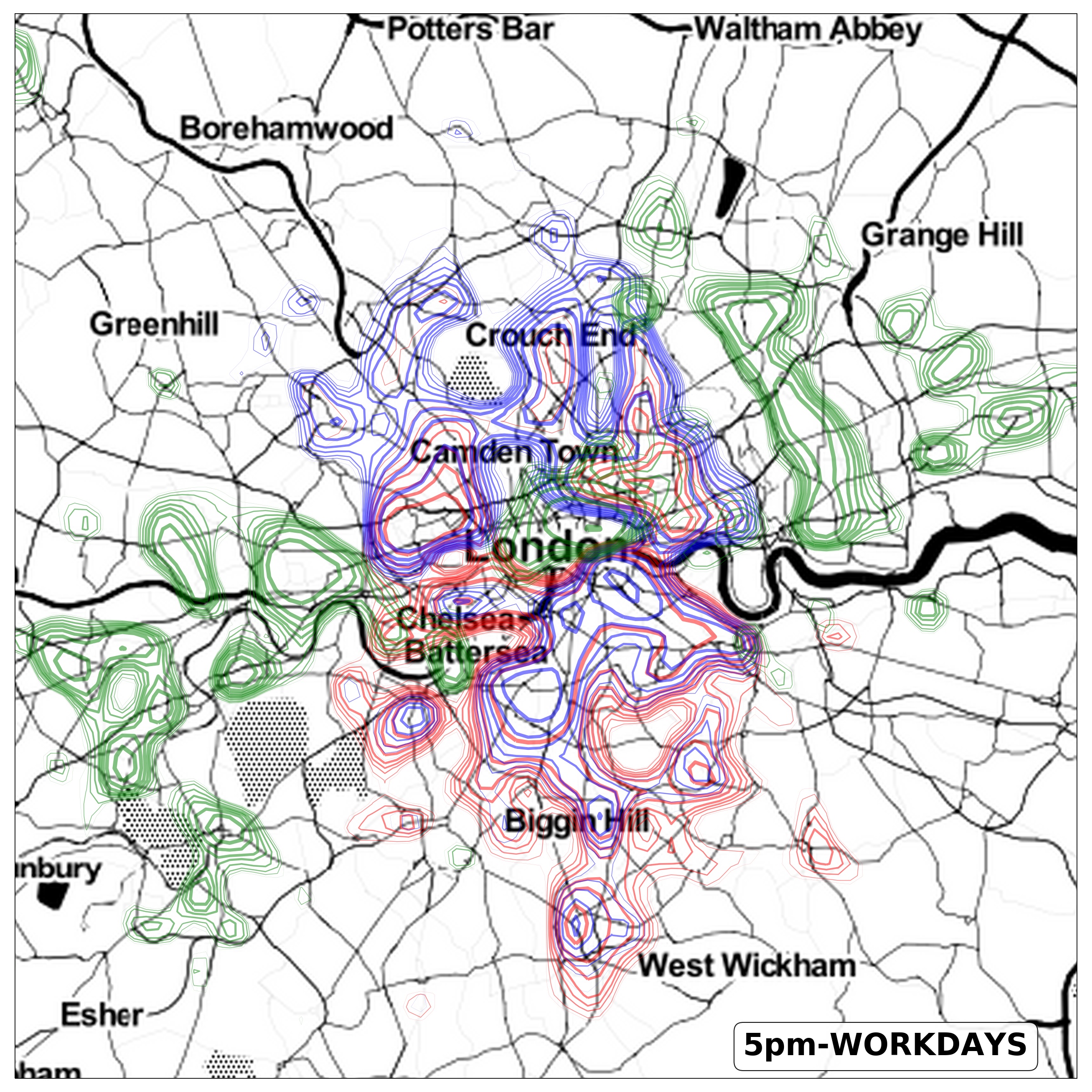}
    \includegraphics[width=0.24\textwidth]{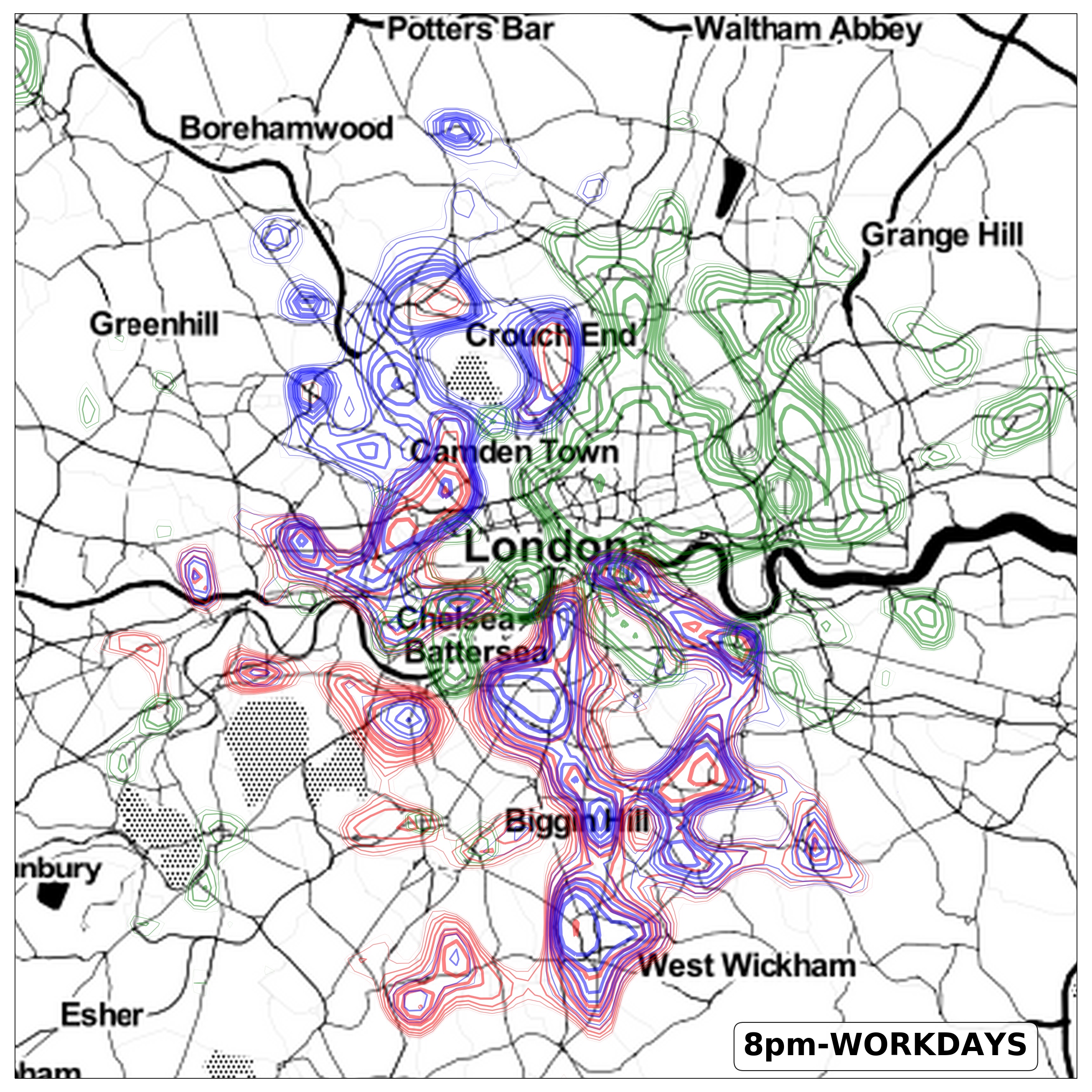}
\caption{Largest (red), second (blue) and smaller clusters ($3rd-1000th$) (green) temporally-averaged spatial distributions over London during the first three months of 2019 at four different hours: 8am, 10am, 5pm, and 8pm. Line width increases with the frequency with which edges belong to each of the main clusters: very thin for $0.85$, thick for $1.00$. A larger version is shown in Fig.\ref{fig:cluster-distrib-weekdays-london}.}
\label{fig:cluster-distrib}
\end{figure*}

Graphically displaying the features of critical configurations at a particular time is a difficult task due to the intrinsic variability of traffic: the largest functional clusters at criticality apparently move all over the map changing their shapes. A better approach is to consider their average spatial distributions, defined as the number of times that each edge in the graph belongs to the largest cluster, to the second, and so on, for all percolation instances of the system over some period, and to bin this information on a grid on top of the city map. We present in Fig. \ref{fig:cluster-distrib} the spatial distributions for the first (red), second (blue) and the union of the rest ($3rd-1000th$ as a whole in green) of the clusters for London during weekdays: the largest component is most frequently seen in the southern part of the city at all hours except between 10:00am and 11:00am (second from left) when it dominates the northern part of the city instead. River Thames, as expected, acts as a natural divider of the urban topology creating a fragile interface across which traffic is easily split at all times of the day, even for non-rush hours. This natural interface with limited transportation capability is often inhabited by smaller clusters at criticality, especially in the central part of the city.
The second cluster appears especially in the north with the same exception as before, and it shows different degrees of sparseness, that reaches a peak at 8pm (rightmost). Minor clusters usually dominate in the very center of the City at all hours, but extend to the west during rush hours and to the east after noon. Each hour shows its own peculiar average structure during the first three months of 2019. What happens during weekends is visible in Fig. \ref{fig:cluster-distrib_weekeends-london}: both morning hours look very similar to the 10am weekday seen before, with the largest cluster firmly in the north, with a clear inversion taking place at 5pm. Evening configurations look very sparse for the largest two clusters almost vanishing from the city center in favor of the minor ones. Eastern areas are dominated at all hours by smaller clusters. April, May and June 2019 confirm the same trend, so it is plausible that this scenario holds during the whole year. The situation for NYC is harder to describe (see Figs. \ref{fig:cluster-distrib-weekdays-nyc} and \ref{fig:cluster-distrib-weekends-nyc}) because the UBER dataset is less spatially homogeneous being truncated along state borders and naturally by the sea (Fig. \ref{fig:map-nyc}), so its graph is very irregular (and sparsely matches the dataset) with respect to London. Nonetheless, the main conclusions standing for London apply to NYC.

Another approach to visualize how breakup patterns evolve is to compute, for each edge, the average and standard deviation of the size of the cluster to which it belongs. In Figs. \ref{fig:size-variance-weekdays-london} and \ref{fig:size-variance-weekends-london}, we plotted the spatial distributions of these two quantities at all hours for weekdays and weekends, respectively, finding recognizable patterns complementing the occurrence distributions of Fig.\ref{fig:cluster-distrib}.


\begin{figure}[ht!]
\centering
    \includegraphics[width=0.47\textwidth]{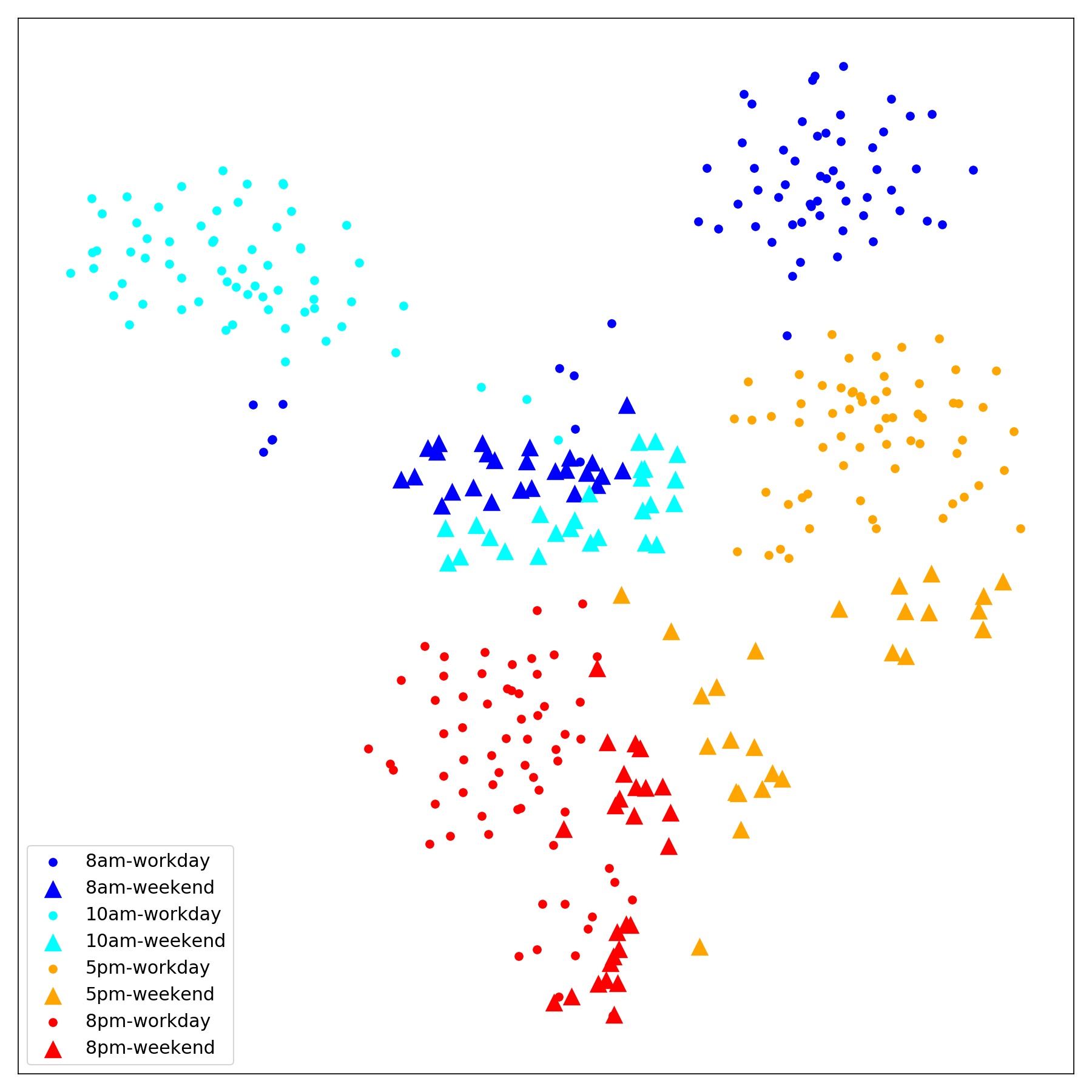}
\caption{Distribution of London traffic configurations over three months obtained from t-SNE and Fowlkes-Mallows similarity: each symbol represents a single day at a specific hour.}
\label{fig:tsne_london}
\end{figure}

After showing the main average properties of the underlying traffic structures during the day, we verify whether single traffic instances possess a detectable fingerprint of their average behavior or not. In other words, we want to estimate the variability of the clusterings at equivalent times.
In Fig.\ref{fig:tsne_london}, we show the 2D representation of the traffic configurations (at criticality) for the first $90$ days of $2019$ at 8am (blue), 10am (cyan), 5pm (orange) and 8pm (red) for the city of London obtained by dimensionality reduction of the Fowlkes-Mallows similarities via t-SNE~\cite{van2008visualizing}. Triangles represent weekends, and small circles weekdays (even if they were holidays, hence a few outliers). It is immediately apparent that configurations relative to the same hours tend to cluster together and that weekdays and weekends, at the same hour, occupy very distinct positions except for 8pm (red circles and triangles) that share roughly the same region. Morning hours (8am in blue and 10am in cyan) on weekends share the central spot, while on weekdays they are very well separated. Overall, weekend configurations tend to create nearby clouds extending from the center towards the lower-right corner of the plot. Rush hours (blue and orange circles) are very distinct, but neighbors, in the top right part of the graph. We verified that the plot is stable against a wide choice of t-SNE parameters~\cite{scikit-learn} (i.e., perplexity and learning-rate). NYC's t-SNE plots are shown in Fig. \ref{fig:t-sne-nyc}, and distinct clusters are recognizable, though less than for London. The results for April, May and June 2019 are qualitatively the same as the three preceding months for both cities. 

Finally, we note that random percolation produced critical configurations clustered together, far from real traffic. On the other hand, spatially-correlated percolation with $\lambda\approx 0.1-0.5$ created network states that t-SNE places very close to morning rush hours between weekdays and weekends (Fig. \ref{fig:t-sne-random-london}).
The same approach used to prove predictability of configurations for different days (even months apart), but equivalent hour, may be used to show that breakup patterns are generally stable if analyzed one hour later, as shown in Fig~\ref{fig:t-sne-london-extended_hours} in which cold and warm shades of color represent morning and afternoon/evening hours, respectively. The only exception is observed during the morning rush hour (8am blue vs 9am dark blue): a symptom of an abrupt global traffic change. Conversely, states belonging to 9, 10 and 11am (cyan, dark cyan and dark blue) create a single large cluster in which the last one is linearly separable from the rest.

{\it Discussion ---} We studied two large, densely populated, metropolitan areas for which an extensive dataset of real velocities exists by means of a percolation approach, both for real data and synthetic traffic with variable spatial correlations. The results strongly prove the existence of an underlying stable structure specific for each time of the day during weekdays and weekends. This finding goes beyond the typical differentiation detected so far between rush and non-rush hours: each moment appears to possess a recognizable specificity in the way in which major functional clusters geographically organize at criticality. The critical percolation transition allows to define a standard framework to compare different states of the system (real and synthetic), by exploiting the sensitivity to edge classification to the same or different clusters of the FM similarity index. The same approach should be easily applicable to different domains as electrical grids and other transportation networks, where critical percolation transitions have been observed. Our results show that real-time city-wide traffic diagnosis is possible by taking just one velocity snapshot by integrating road sensors and GPS data from several sources. Finally, we found that spatially-correlated random percolation reproduces some properties typical of real traffic during rush hours: this could be useful when trying to identify the best possible improvements to the urban network and several cheap evaluations are needed.
\\\\
{\it Acknowledgements ---} This work is dedicated to the late Gianluigi Zanetti, who contributed with useful and pleasant discussions, and encouraged us to explore this research field. We wish to thank Francesco Versaci and Enrico Gobbetti for useful discussions. This work has been partially supported by the DIFRA Project and by the TDM Project, both funded by the Regional Authorities of Sardinia.

\bibliography{paper}

\clearpage
\pagebreak
\widetext
\begin{center}
\textbf{\large Supplemental Materials: On the stability of traffic breakup patterns in urban networks}
\end{center}
\setcounter{equation}{0}
\setcounter{figure}{0}
\setcounter{table}{0}
\setcounter{page}{1}
\makeatletter
\renewcommand{\theequation}{S\arabic{equation}}
\renewcommand{\thefigure}{S\arabic{figure}}
\renewcommand{\bibnumfmt}[1]{[S#1]}
\renewcommand{\citenumfont}[1]{S#1}
\renewcommand{\thetable}{S\arabic{table}}

\section*{Cluster Occurrence Distributions} \label{suppl}

\begin{figure*}[h]
\centering
    \includegraphics[width=0.4\textwidth]{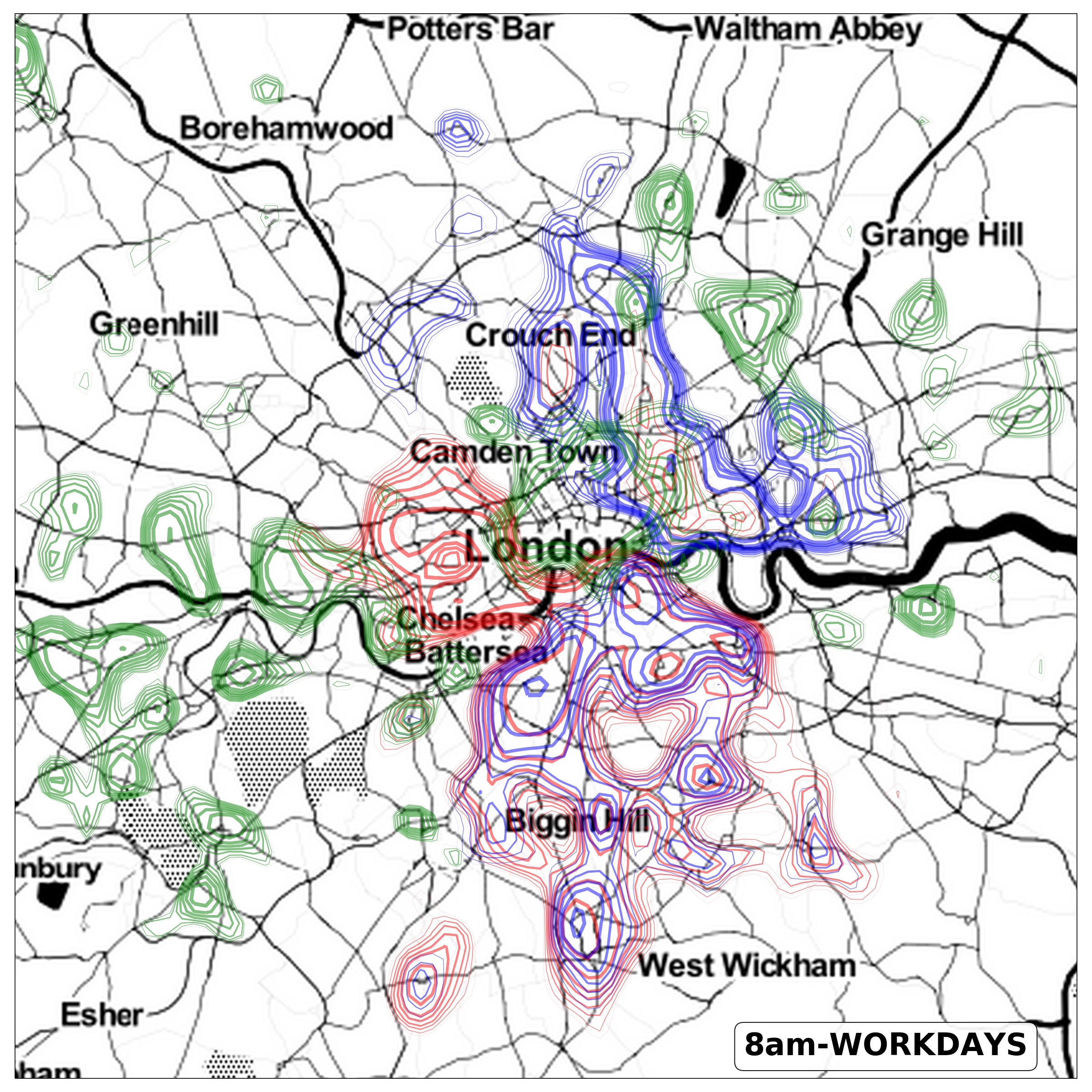}
    \includegraphics[width=0.4\textwidth]{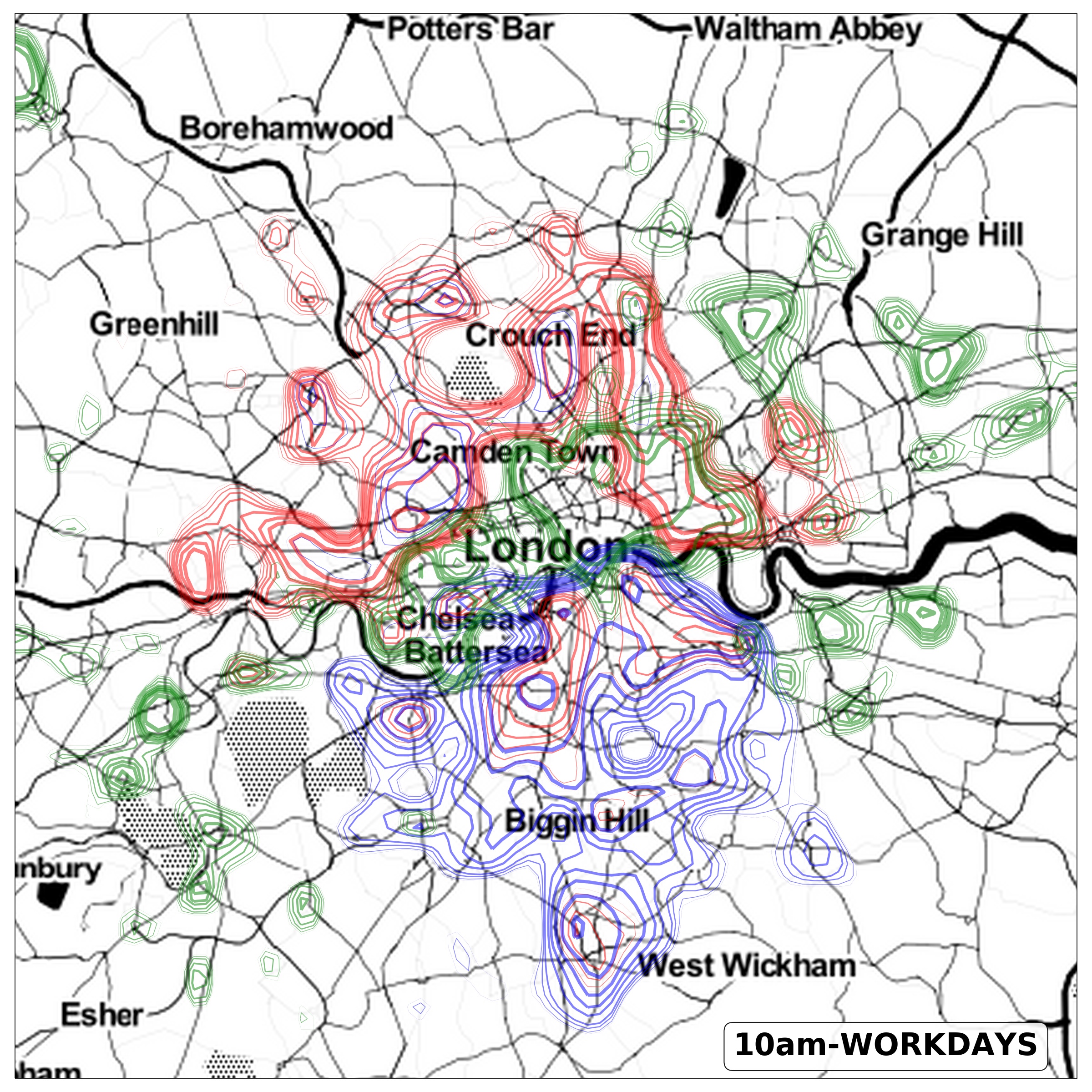}
    \includegraphics[width=0.4\textwidth]{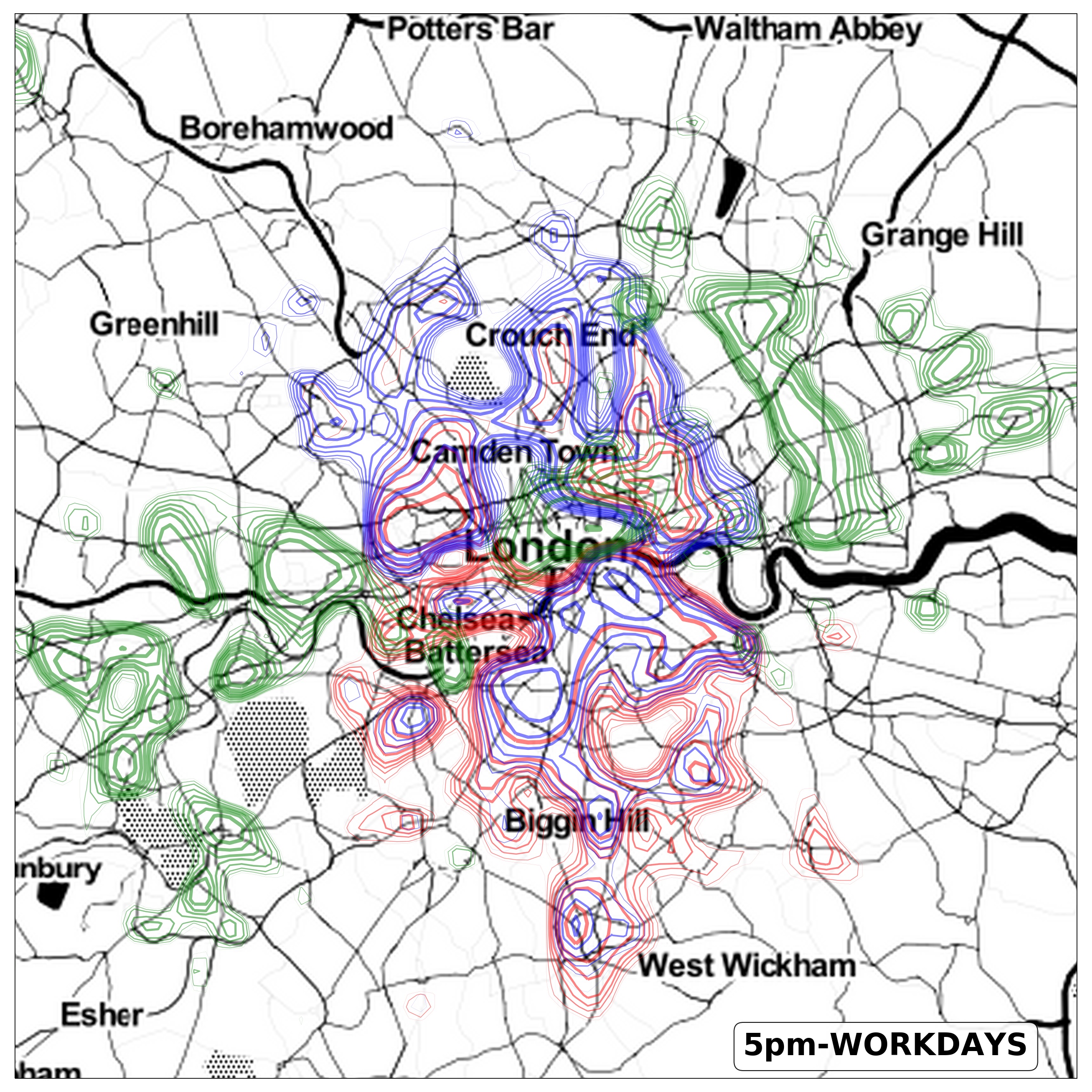}
    \includegraphics[width=0.4\textwidth]{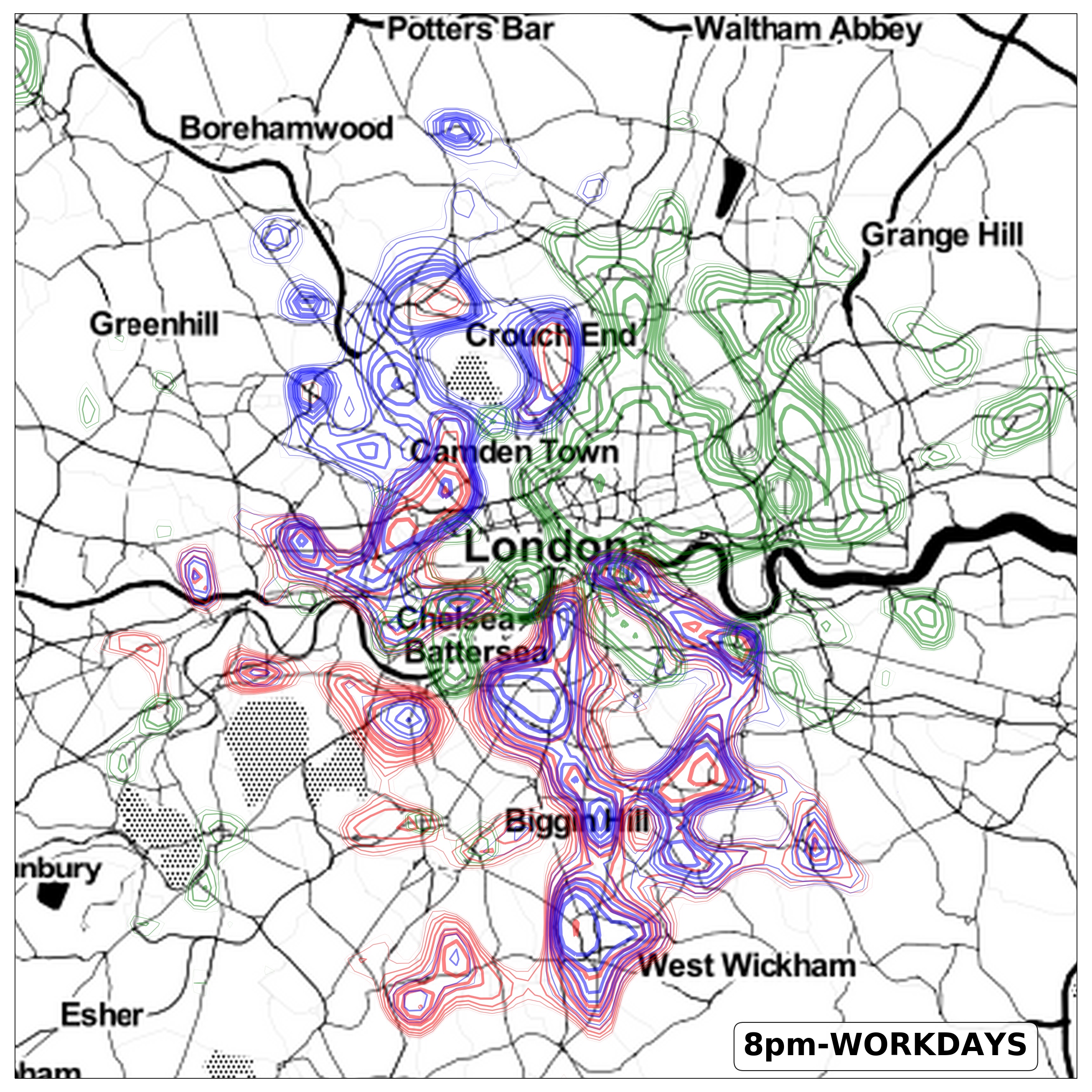}
\caption{Weekdays: Largest (red), second (blue) and smaller clusters ($3rd-1000th$) (green) temporally-averaged spatial distributions over London during the first three months of 2019 at four different hours: 8am, 10am, 5pm, and 8pm. Line width increases with the frequency with which edges belong to each of the main clusters: starting from $0.85$ to $1.00$.}
\label{fig:cluster-distrib-weekdays-london}
\end{figure*}

\begin{figure*}[h]
\centering
    \includegraphics[width=0.4\textwidth]{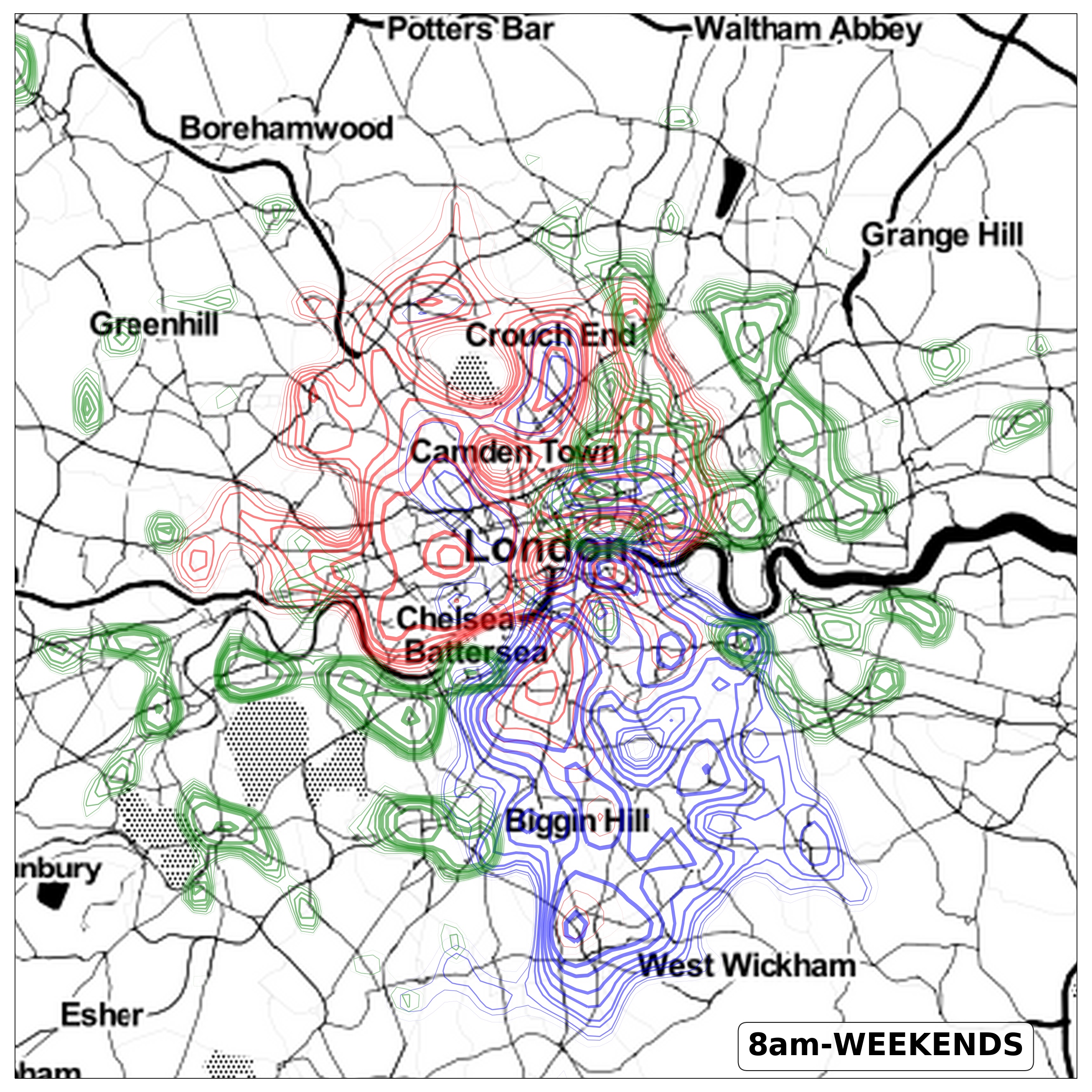}
    \includegraphics[width=0.4\textwidth]{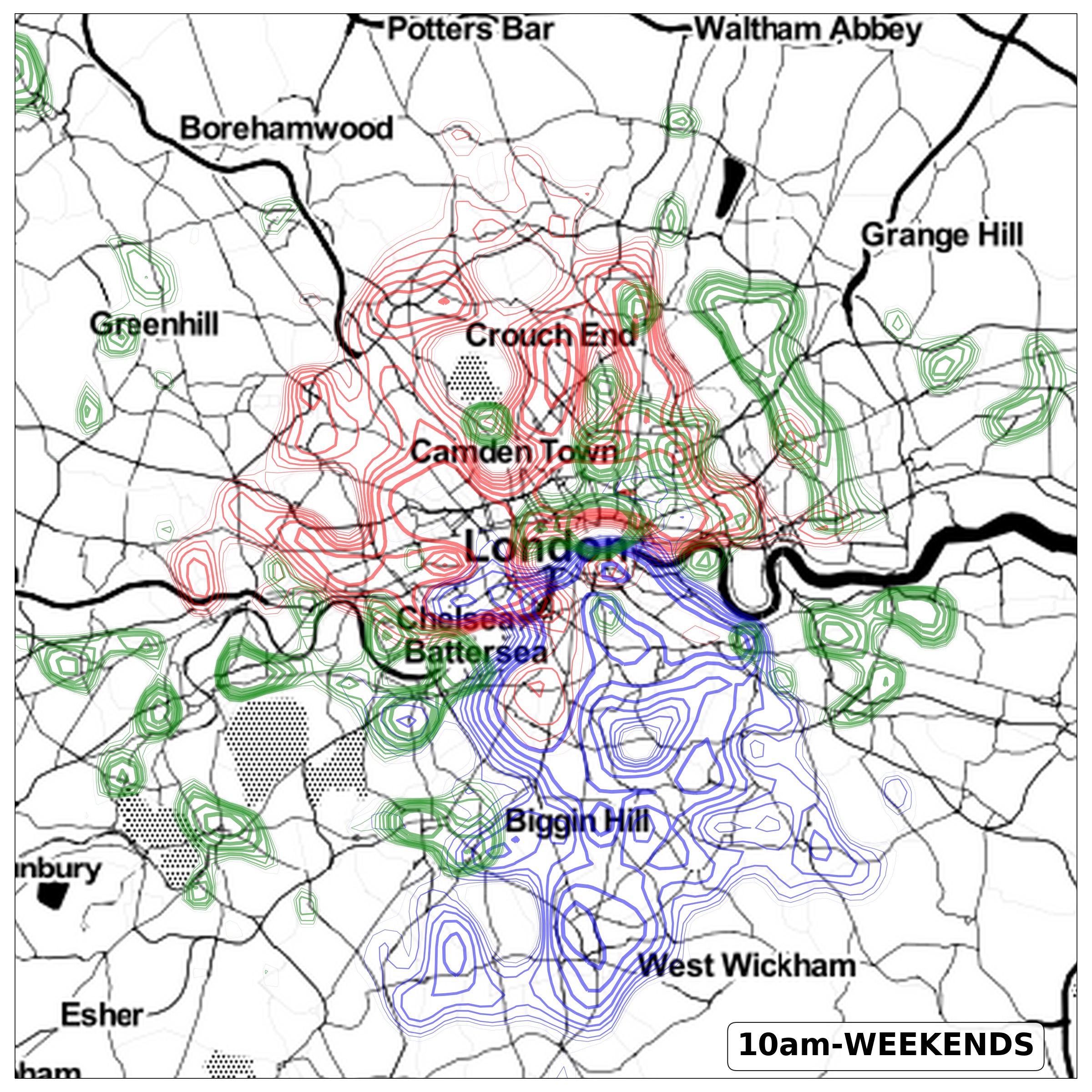}
    \includegraphics[width=0.4\textwidth]{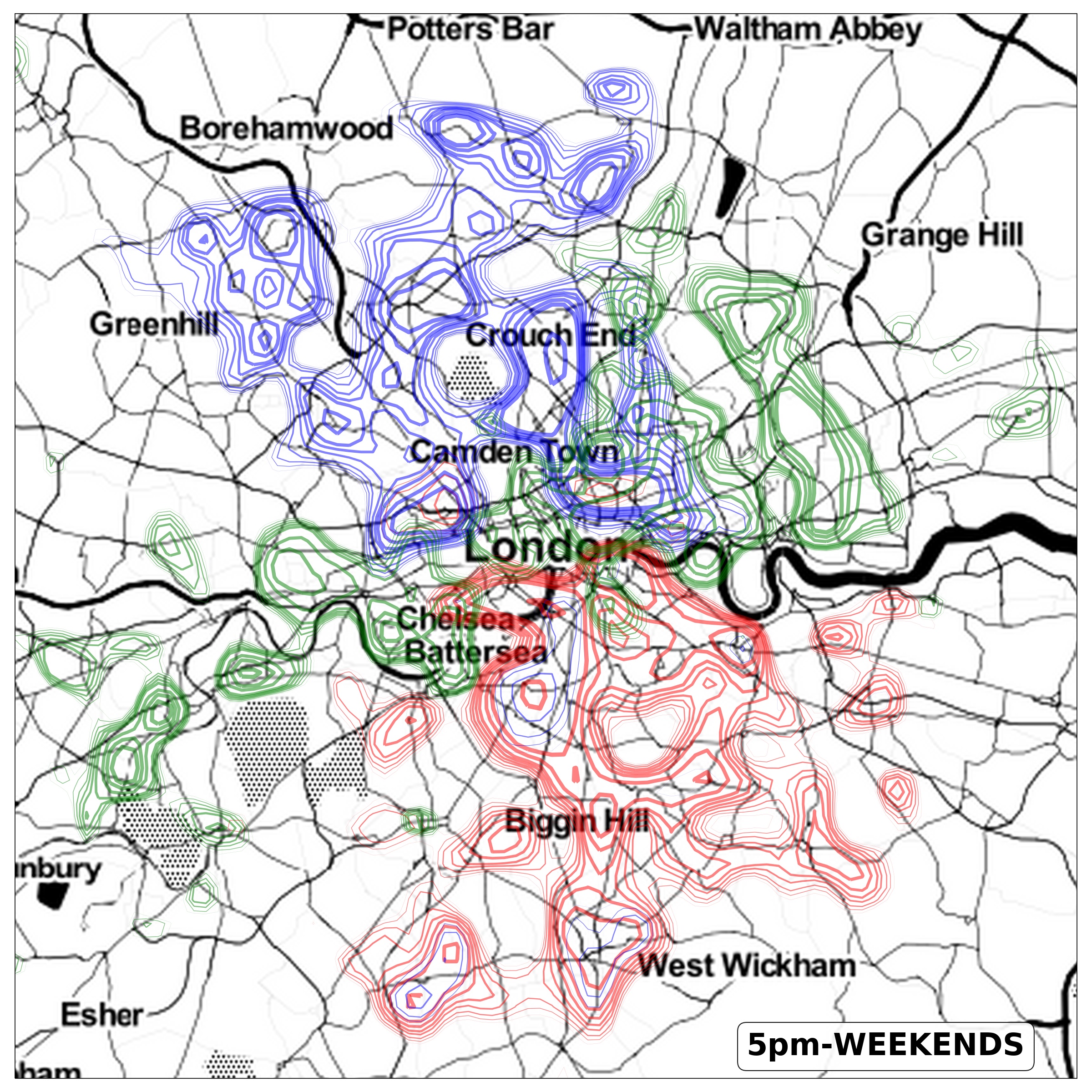}
    \includegraphics[width=0.4\textwidth]{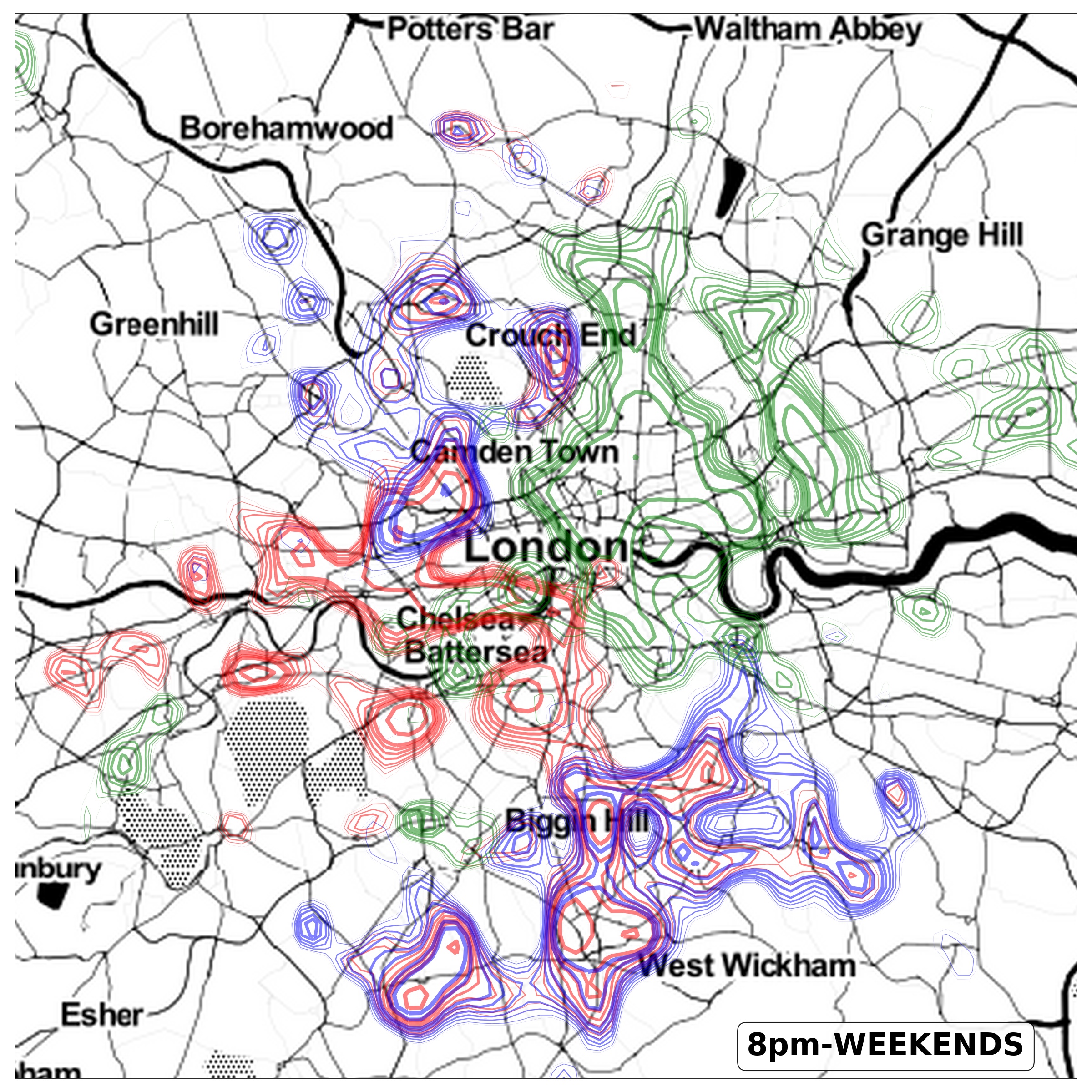}
\caption{Weekends: Largest (red), second (blue) and smaller clusters ($3rd-1000th$) (green) temporally-averaged spatial distributions over London during the first three months of 2019 at four different hours: 8am, 10am, 5pm, and 8pm. Line width increases with the frequency with which edges belong to each of the main clusters: starting from $0.85$ to $1.00$.}
\label{fig:cluster-distrib_weekeends-london}
\end{figure*}

\begin{figure*}[h]
\centering
    \includegraphics[width=0.4\textwidth]{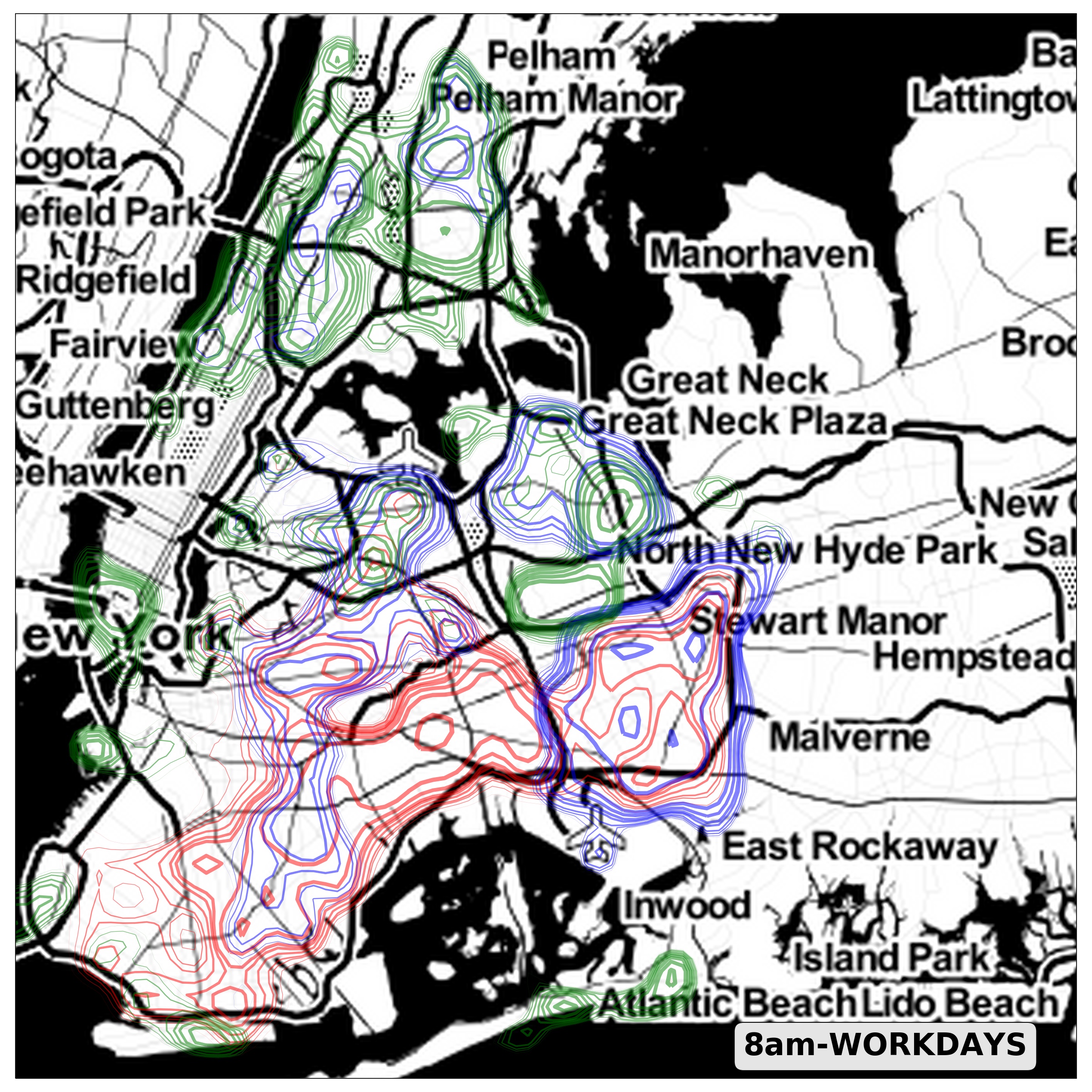}
    \includegraphics[width=0.4\textwidth]{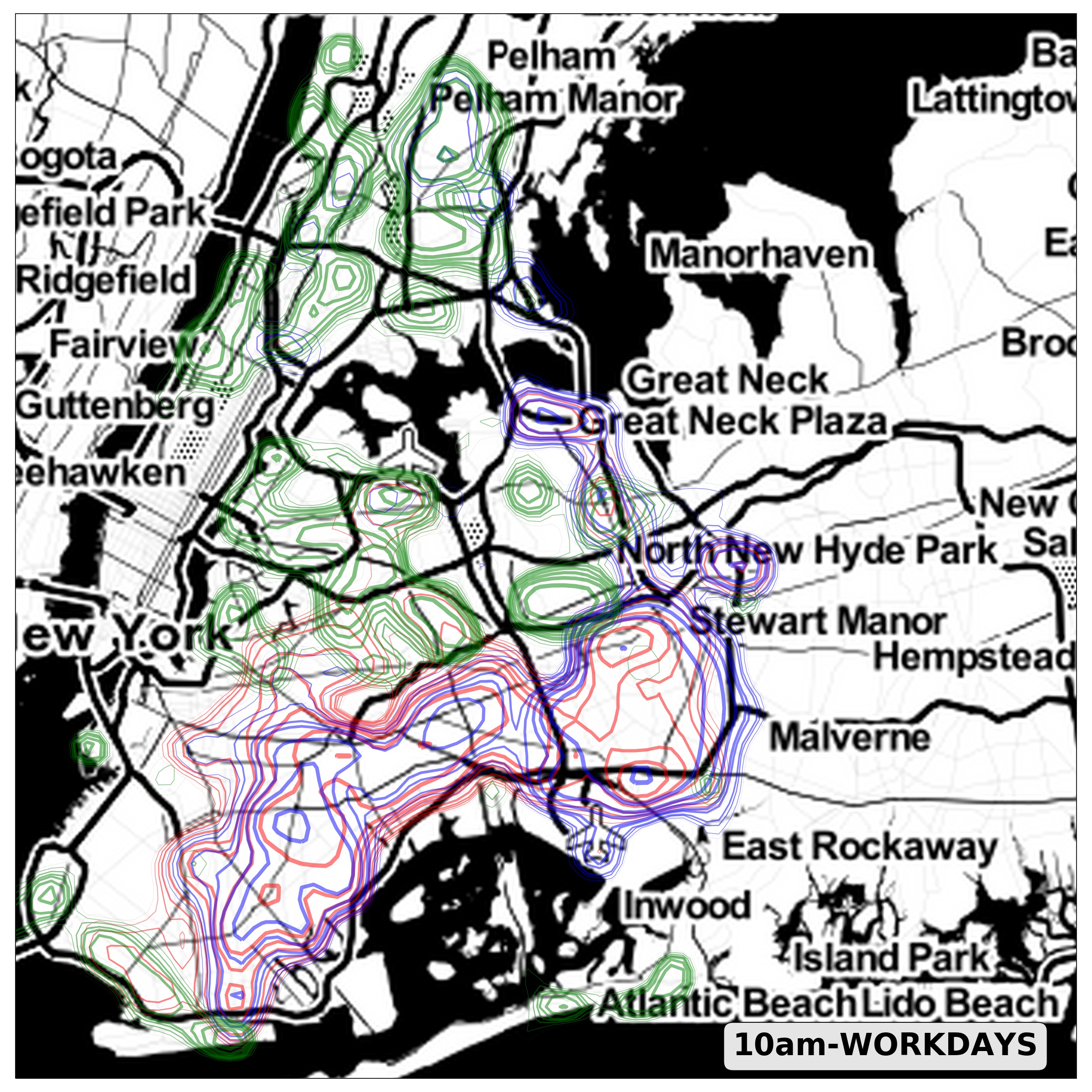}
    \includegraphics[width=0.4\textwidth]{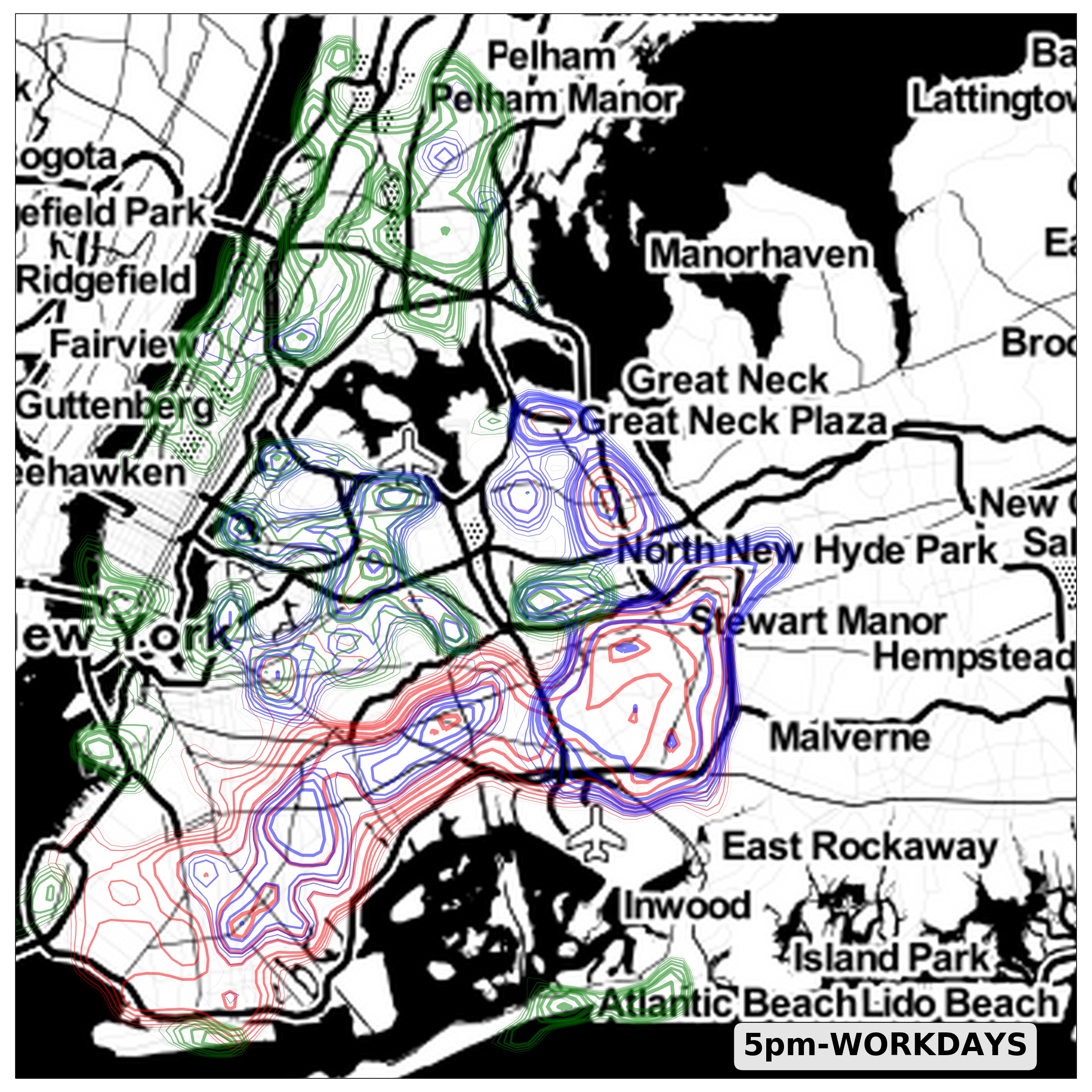}
    \includegraphics[width=0.4\textwidth]{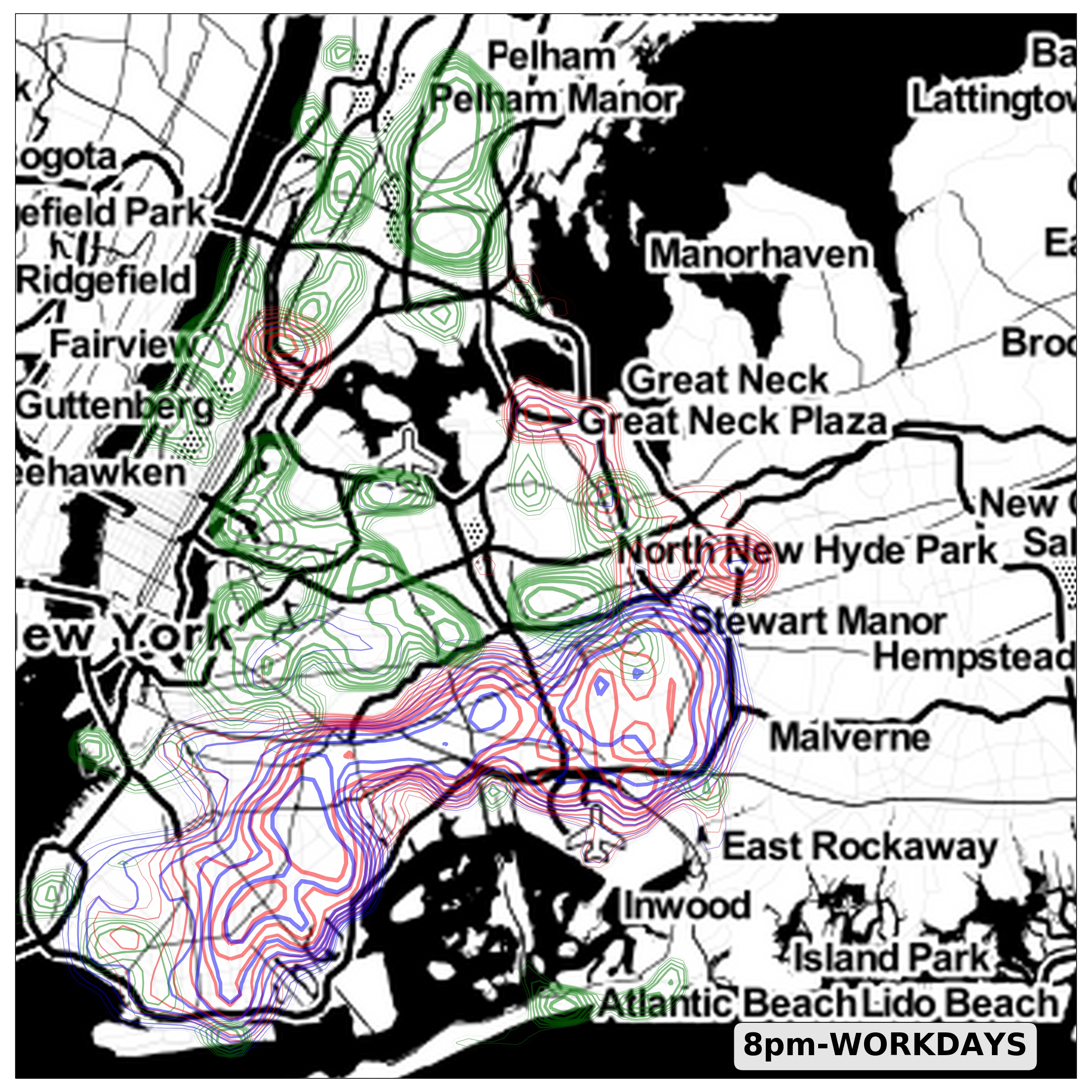}
\caption{Weekdays: Largest (red), second (blue) and smaller clusters ($3rd-1000th$) (green) temporally-averaged spatial distributions over NYC during the first three months of 2019 at four different hours: 8am, 10am, 5pm, and 8pm. Line width increases with the frequency with which edges belong to each of the main clusters: starting from $0.85$ to $1.00$.}
\label{fig:cluster-distrib-weekdays-nyc}
\end{figure*}

\begin{figure*}[h]
\centering
    \includegraphics[width=0.4\textwidth]{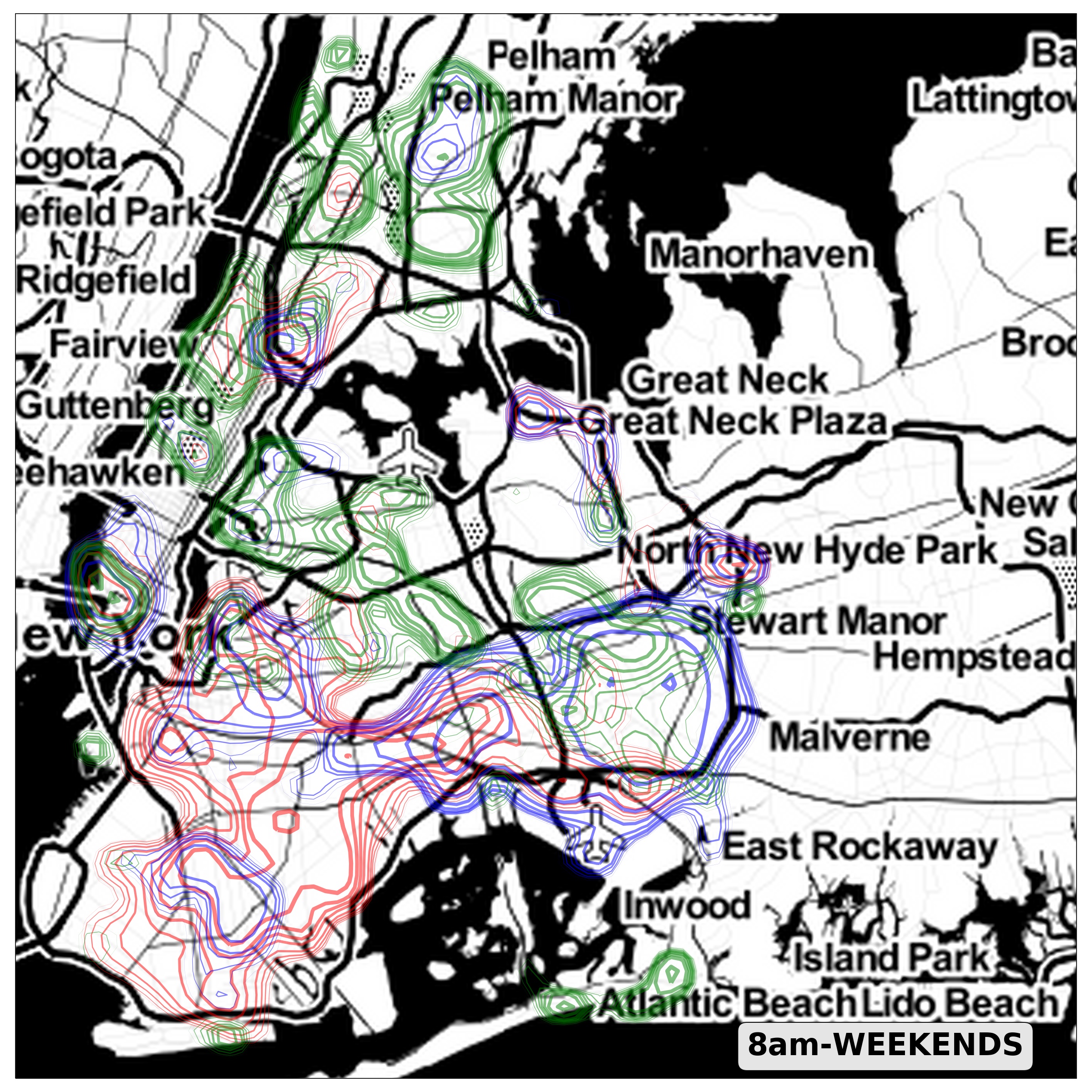}
    \includegraphics[width=0.4\textwidth]{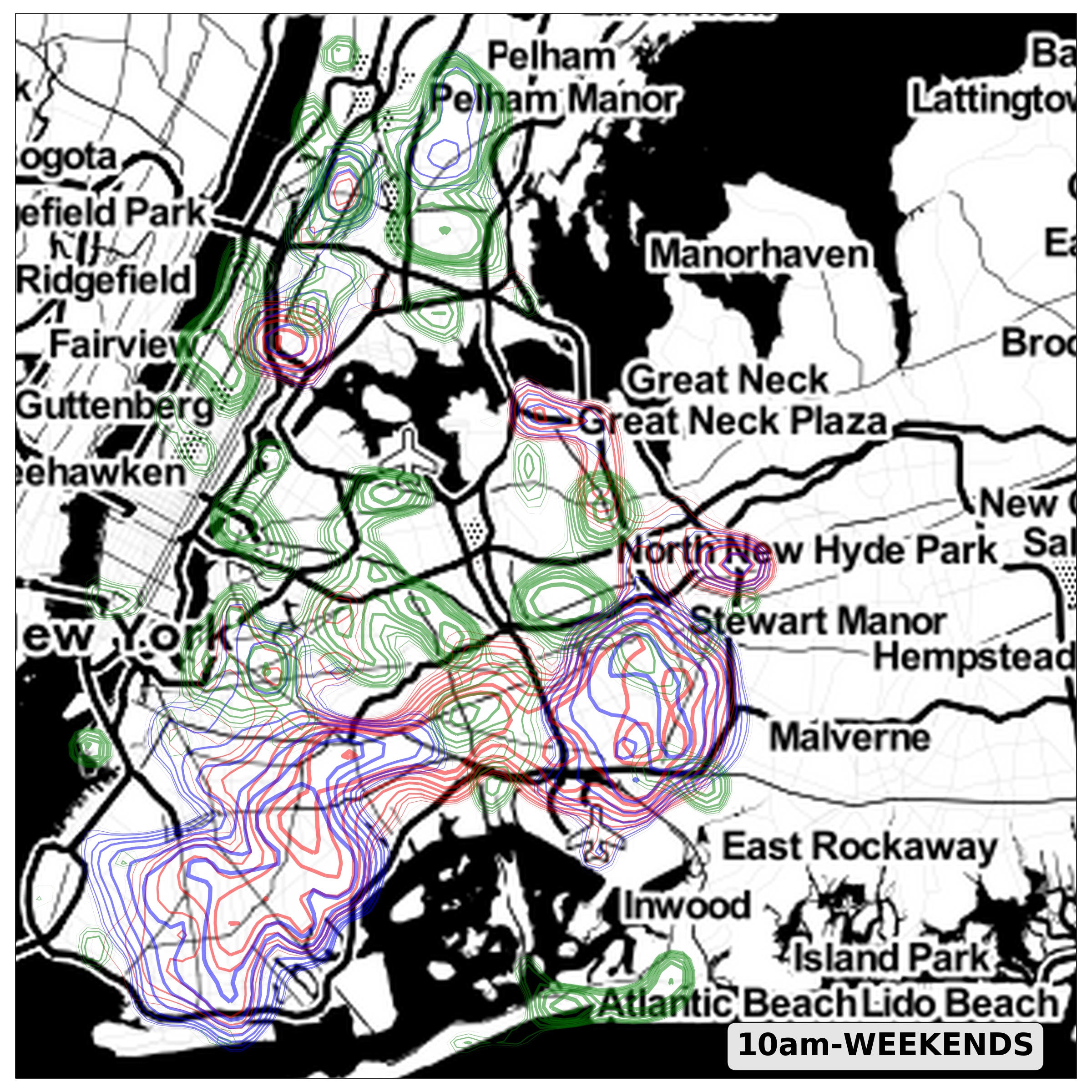}
    \includegraphics[width=0.4\textwidth]{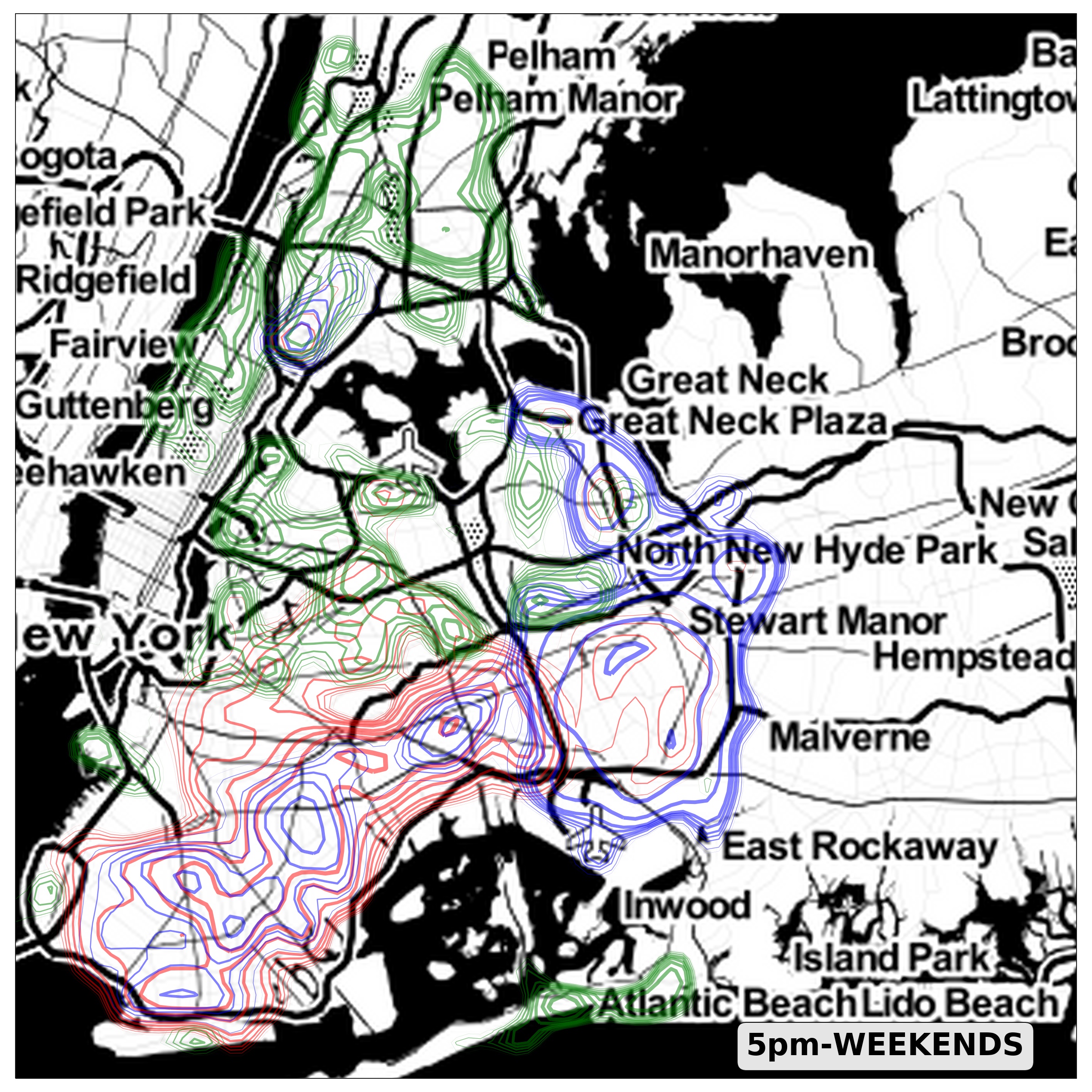}
    \includegraphics[width=0.4\textwidth]{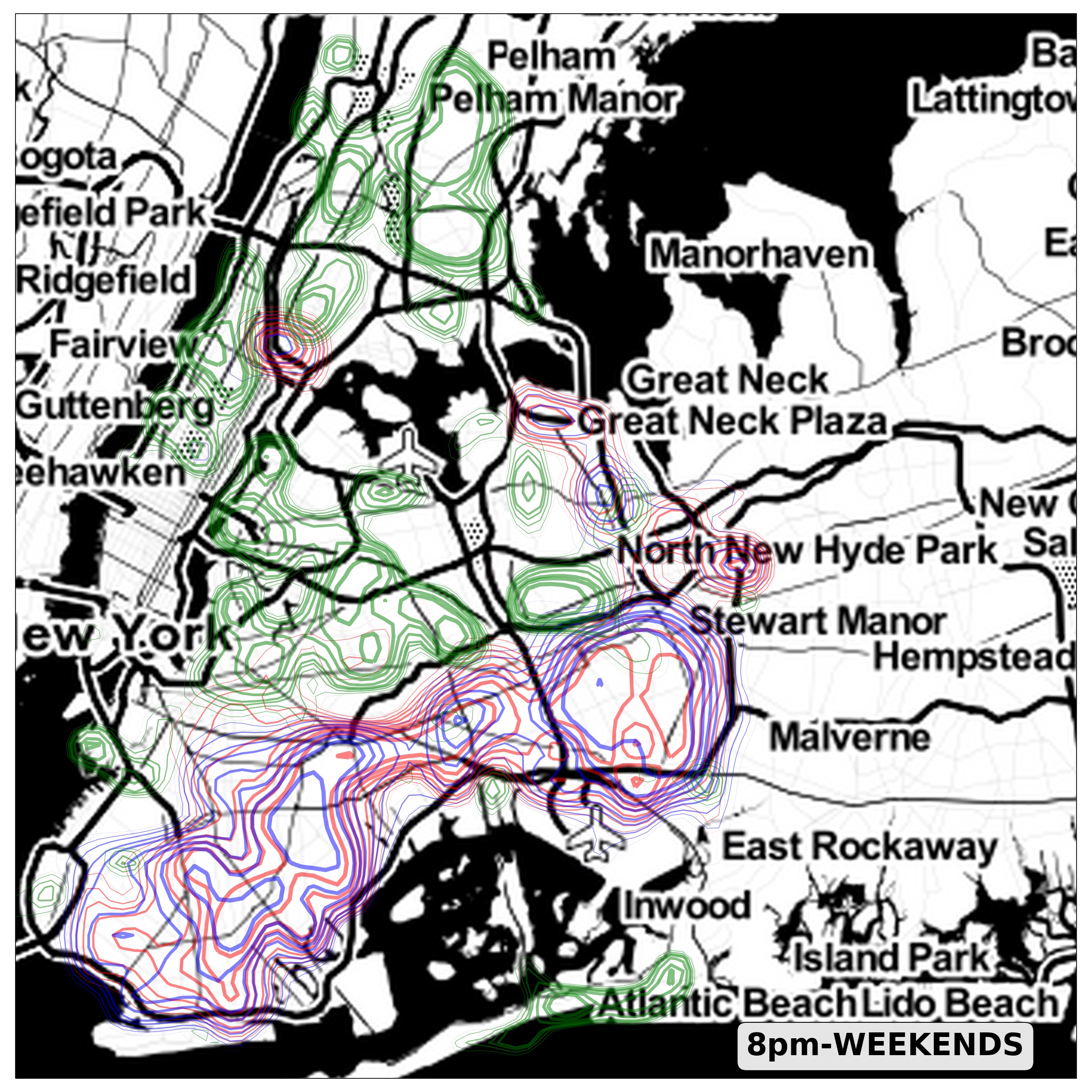}
\caption{Weekends: Largest (red), second (blue) and smaller clusters ($3rd-1000th$) (green) temporally-averaged spatial distributions over NYC during the first three months of 2019 at four different hours: 8am, 10am, 5pm, and 8pm. Line width increases with the frequency with which edges belong to each of the main clusters: starting from $0.85$ to $1.00$.}
\label{fig:cluster-distrib-weekends-nyc}
\end{figure*}

\section*{t-SNE Plots for Fowlkes-Mallows Similarities}

\begin{figure}[H]
\centering
    \includegraphics[width=0.8\textwidth, trim={0 0 0 0},clip]{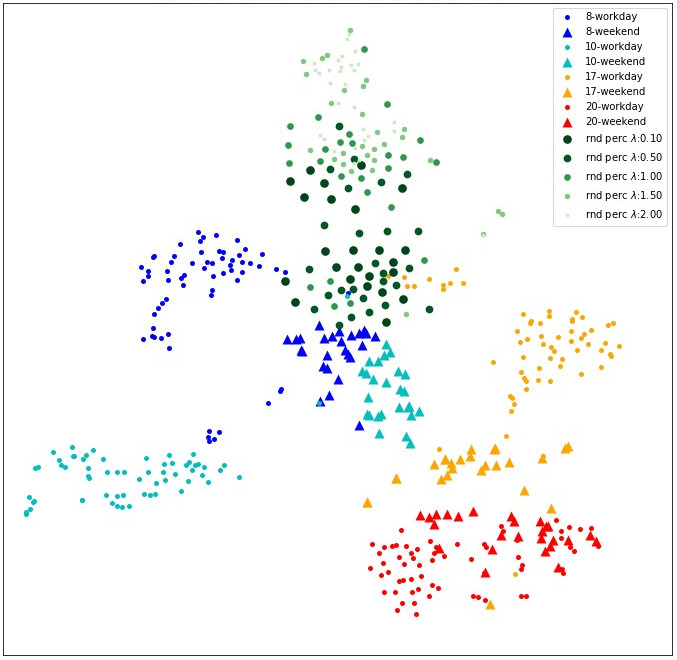}
\caption{Critical percolation for real and synthetic traffic: distribution of London traffic configurations over three months obtained from t-SNE and Fowlkes-Mallows similarity: each symbol represents a single day at a specific hour. Triangles are weekends. Green circles of growing size and darker shade are produced by increasingly correlated random percolation.}
\label{fig:t-sne-random-london}
\end{figure}

\begin{figure}[H]
\centering
    \includegraphics[width=0.8\textwidth, trim={0 0 0 0},clip]{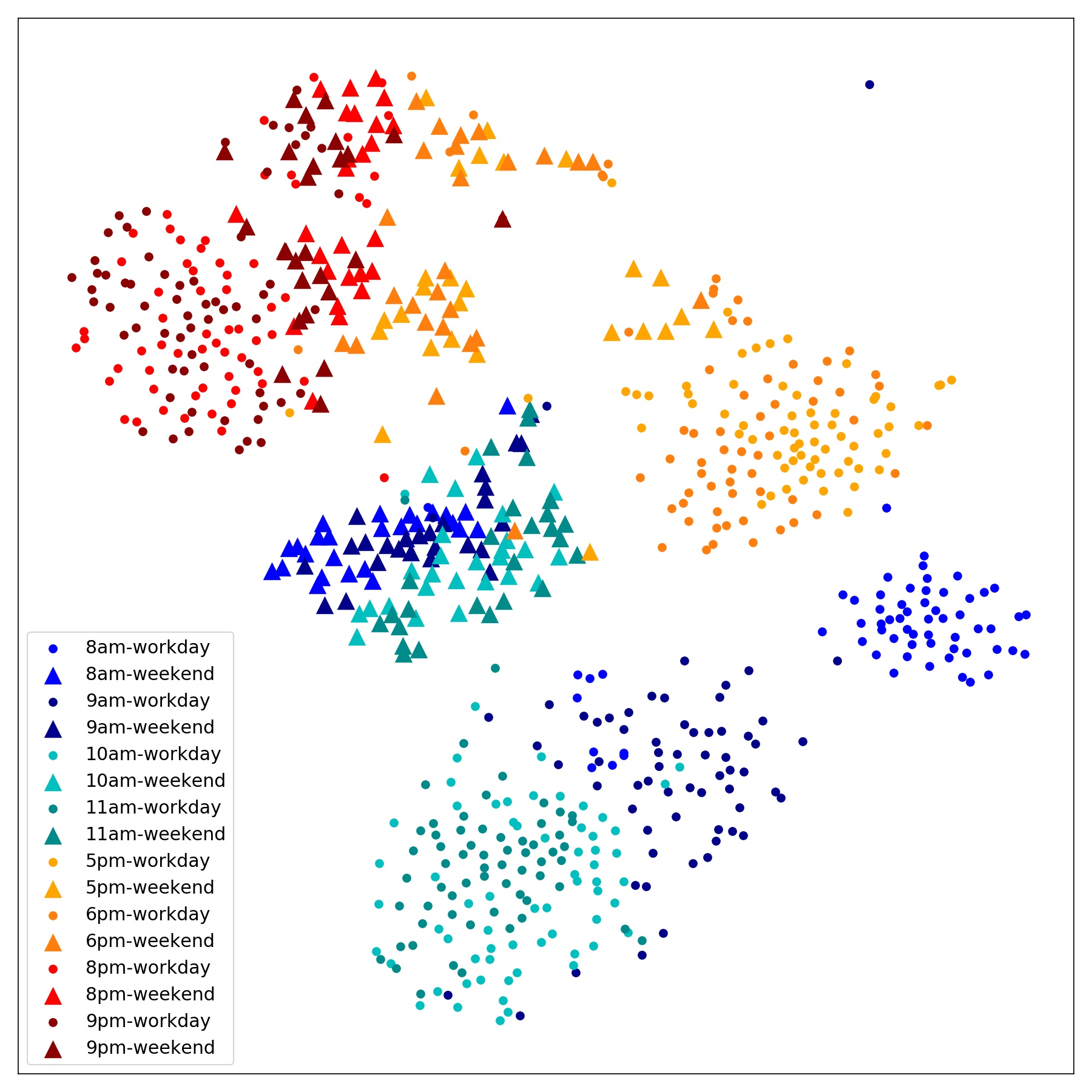}
\caption{Critical percolation for real traffic. Distribution of London traffic configurations over three months obtained from t-SNE and Fowlkes-Mallows similarity. Extended set of times: 8am, 9am, 10am, 11am, 5pm, 6pm, 7pm, 8pm, 9pm. Each symbol represents a single day at a specific hour. Triangles are weekends.}
\label{fig:t-sne-london-extended_hours}
\end{figure}

\begin{figure}[H]
\centering
    \includegraphics[width=0.9\textwidth, trim={0 0 0 0}, clip]{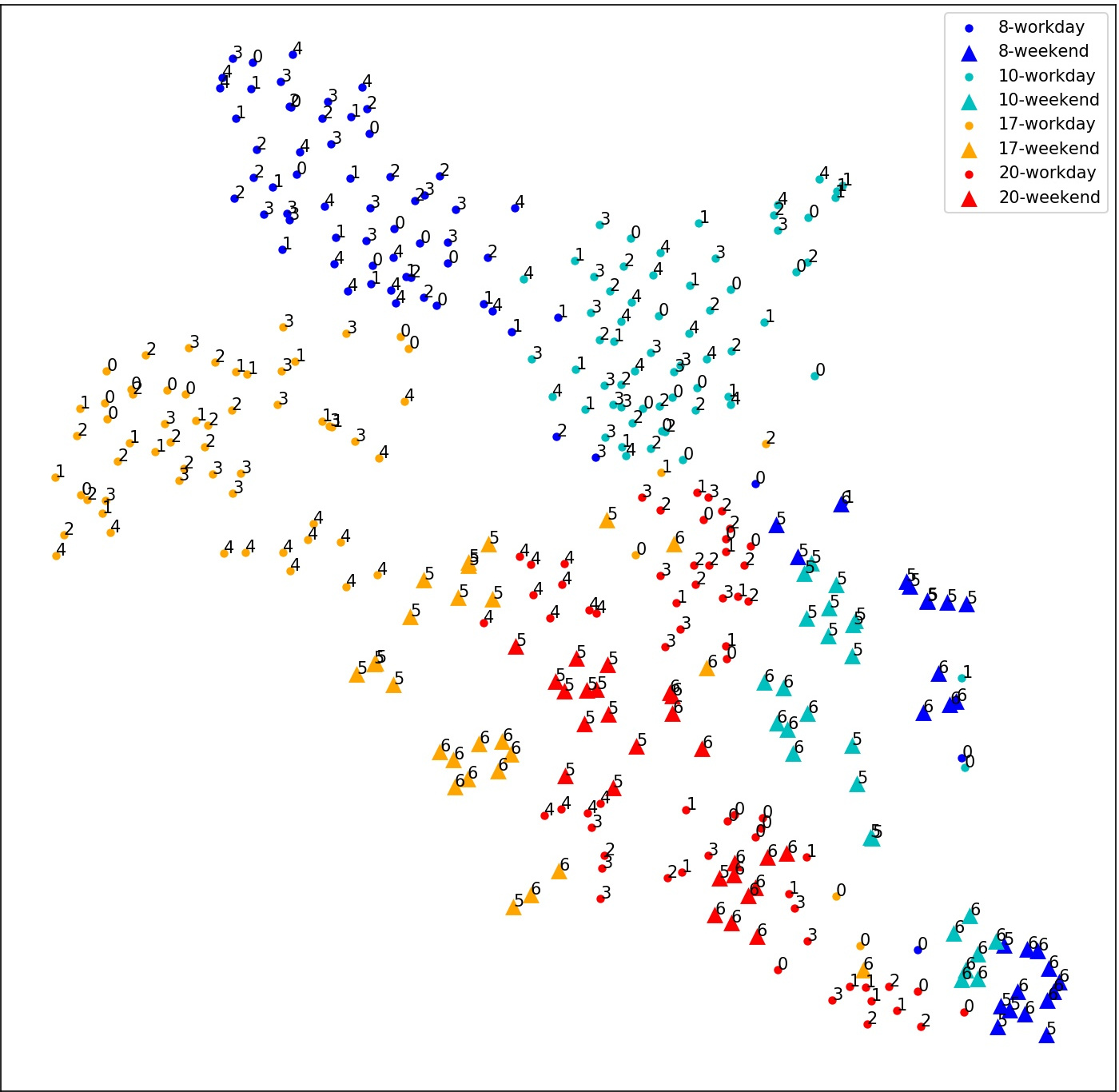}
\caption{Critical percolation: distribution of New York City traffic configurations over three months obtained from t-SNE and Fowlkes-Mallows similarity: each symbol represents a single day at a specific hour. Triangles are weekends, numbers are days of the week starting with Monday.}
\label{fig:t-sne-nyc}
\end{figure}

\section*{Spatial Distributions of the Cluster Size and Standard Deviation and their Relationship}

\begin{figure}[H]
\centering
    \includegraphics[width=0.5\textwidth]{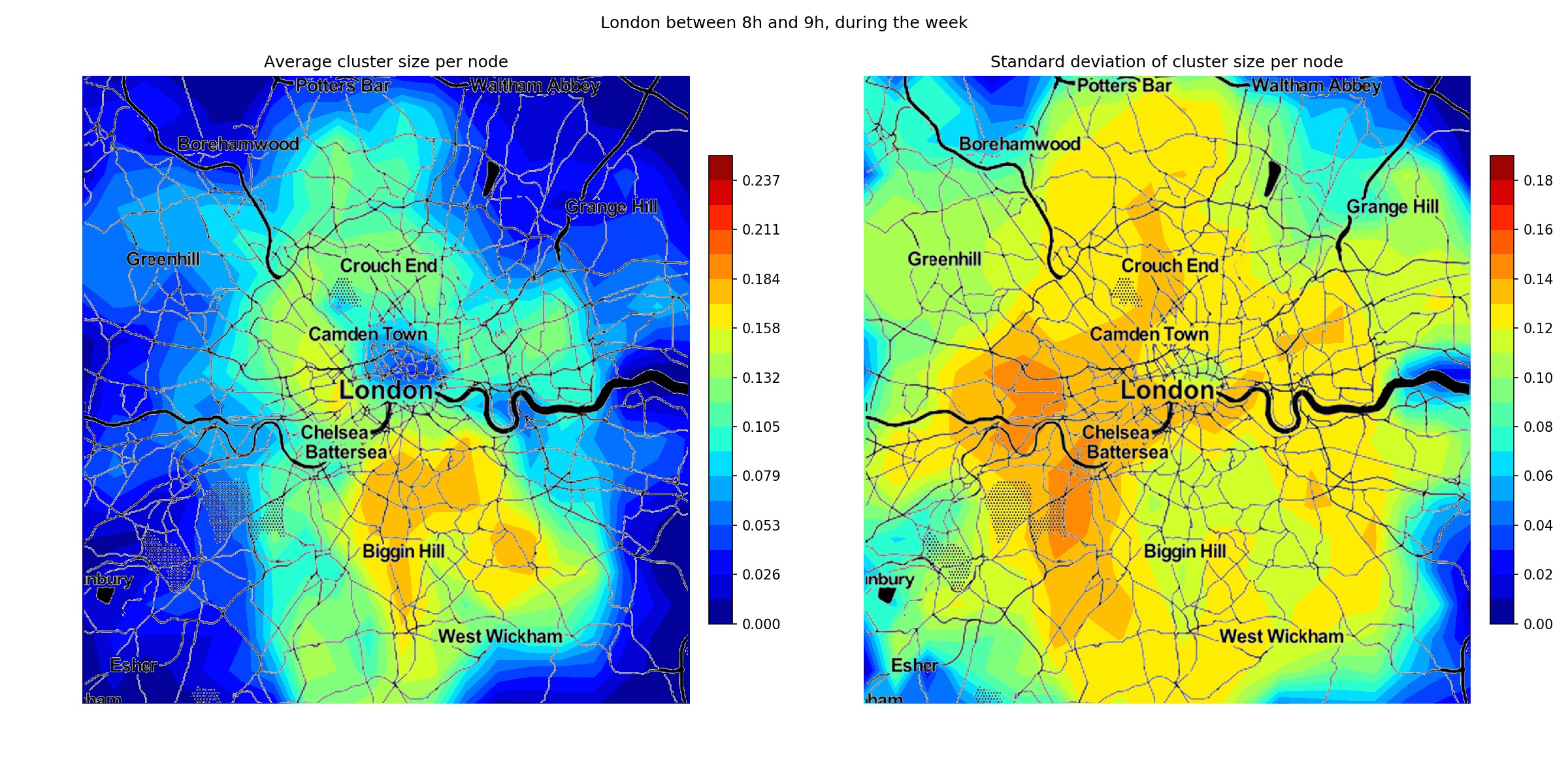}
    \includegraphics[width=0.43\textwidth]{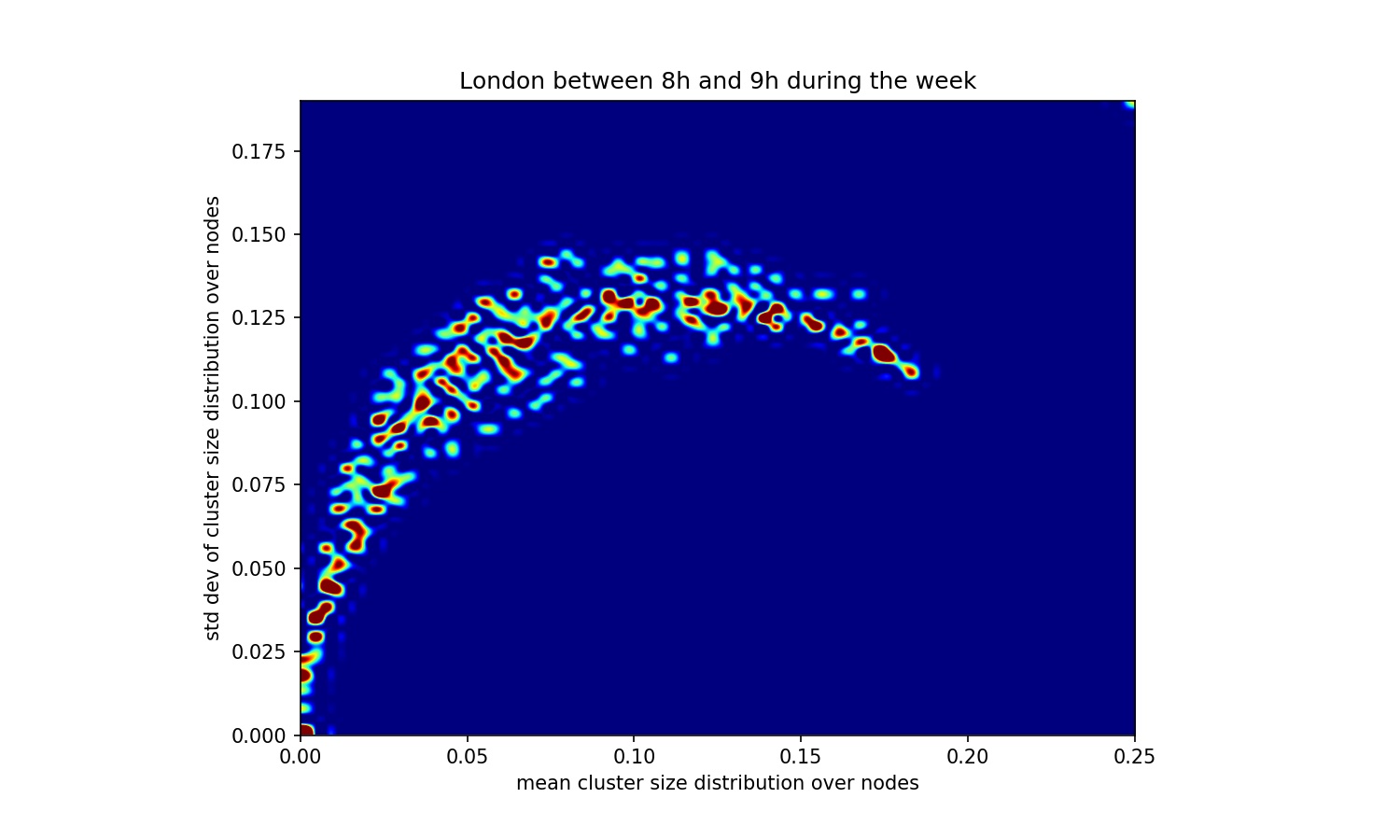}
    \includegraphics[width=0.5\textwidth]{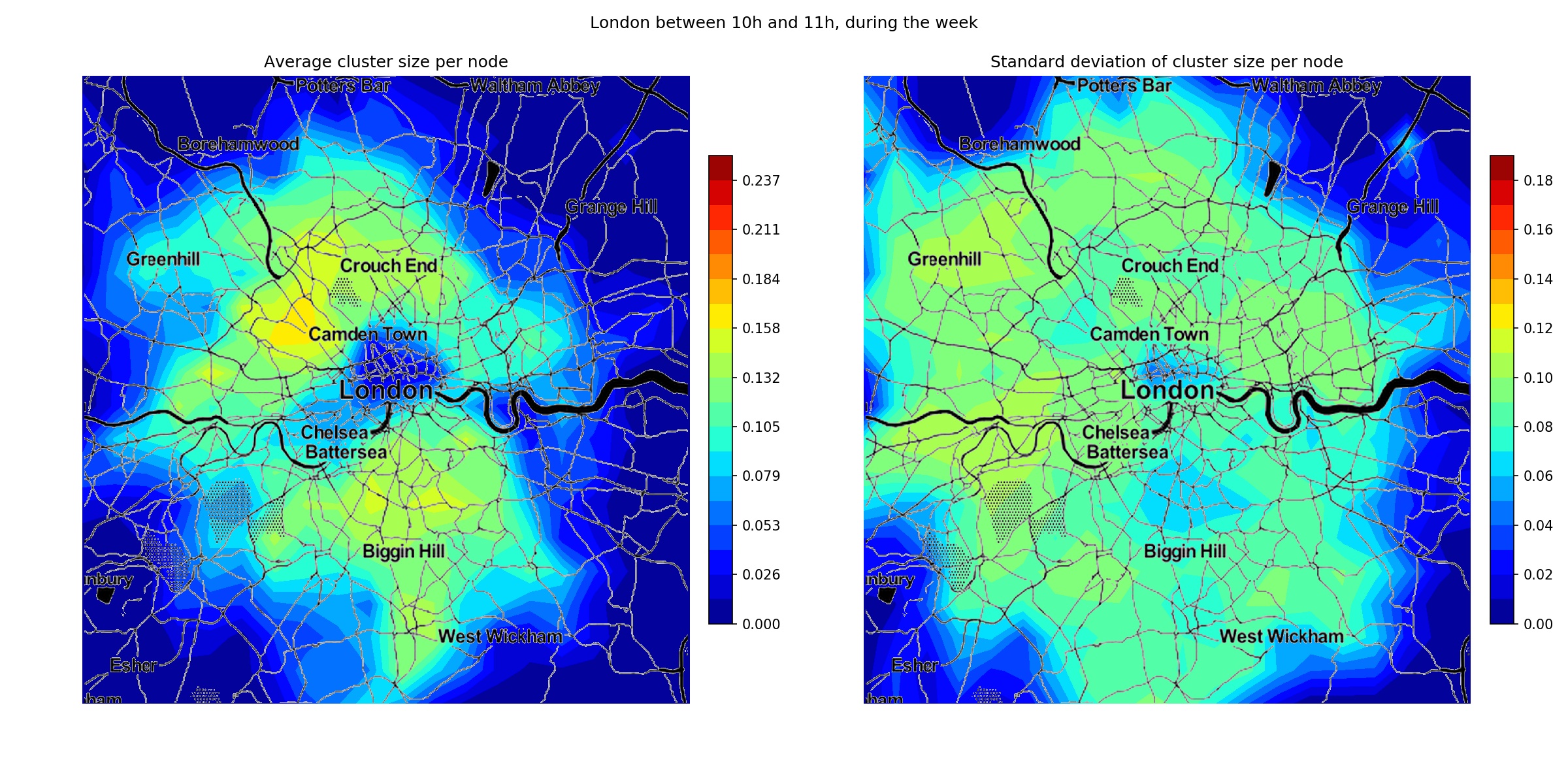}
    \includegraphics[width=0.43\textwidth]{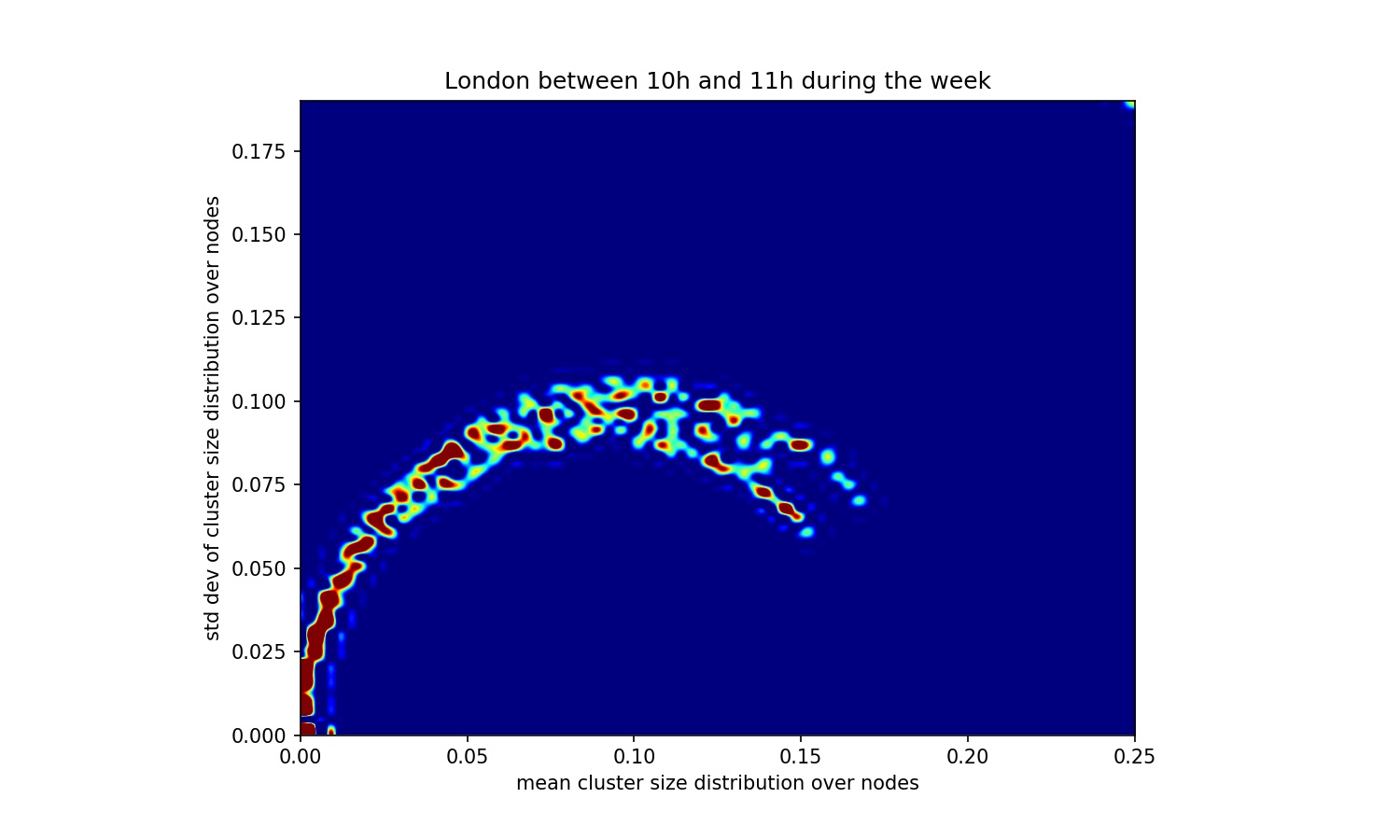}
    \includegraphics[width=0.5\textwidth]{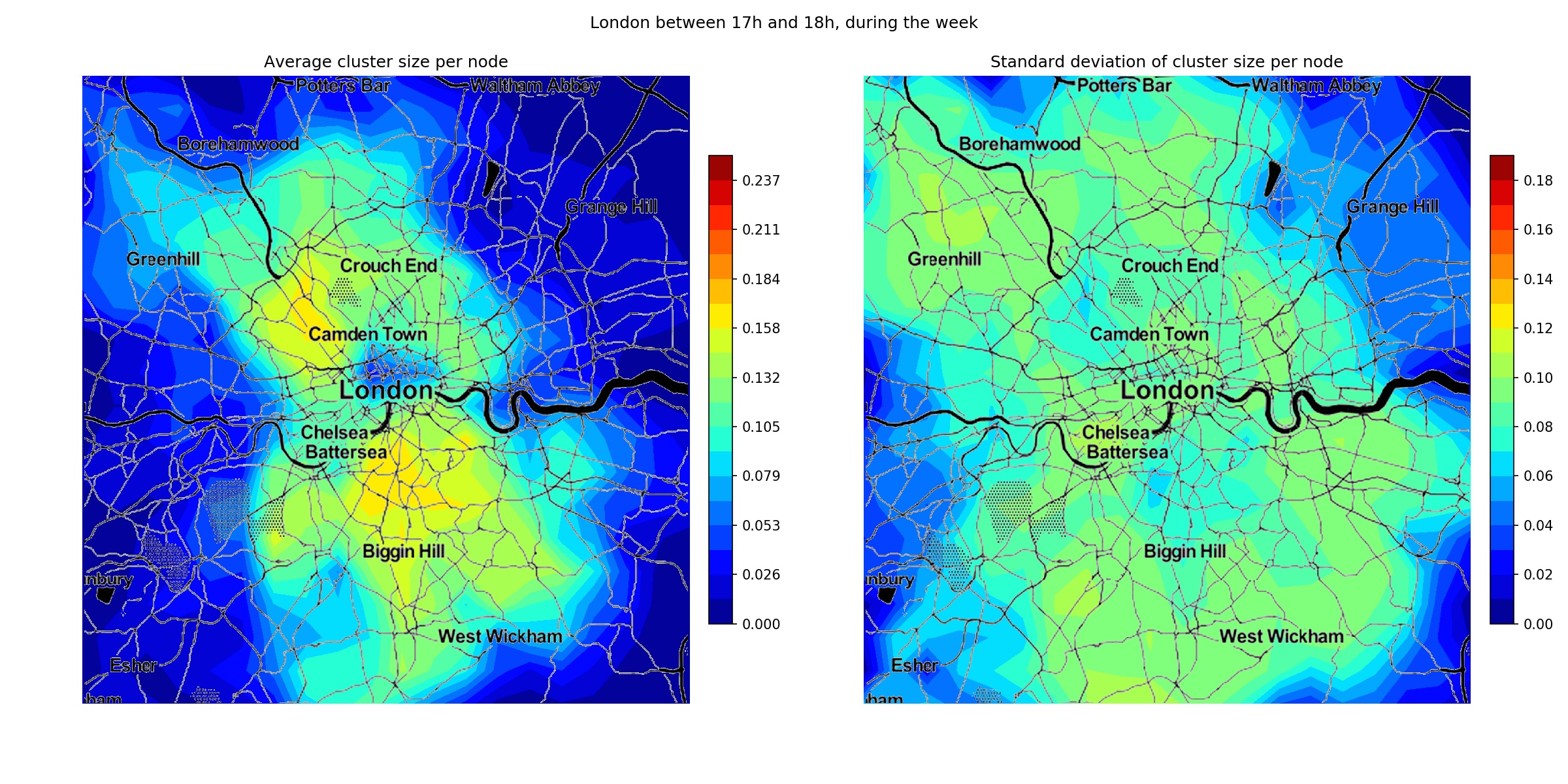}
    \includegraphics[width=0.43\textwidth]{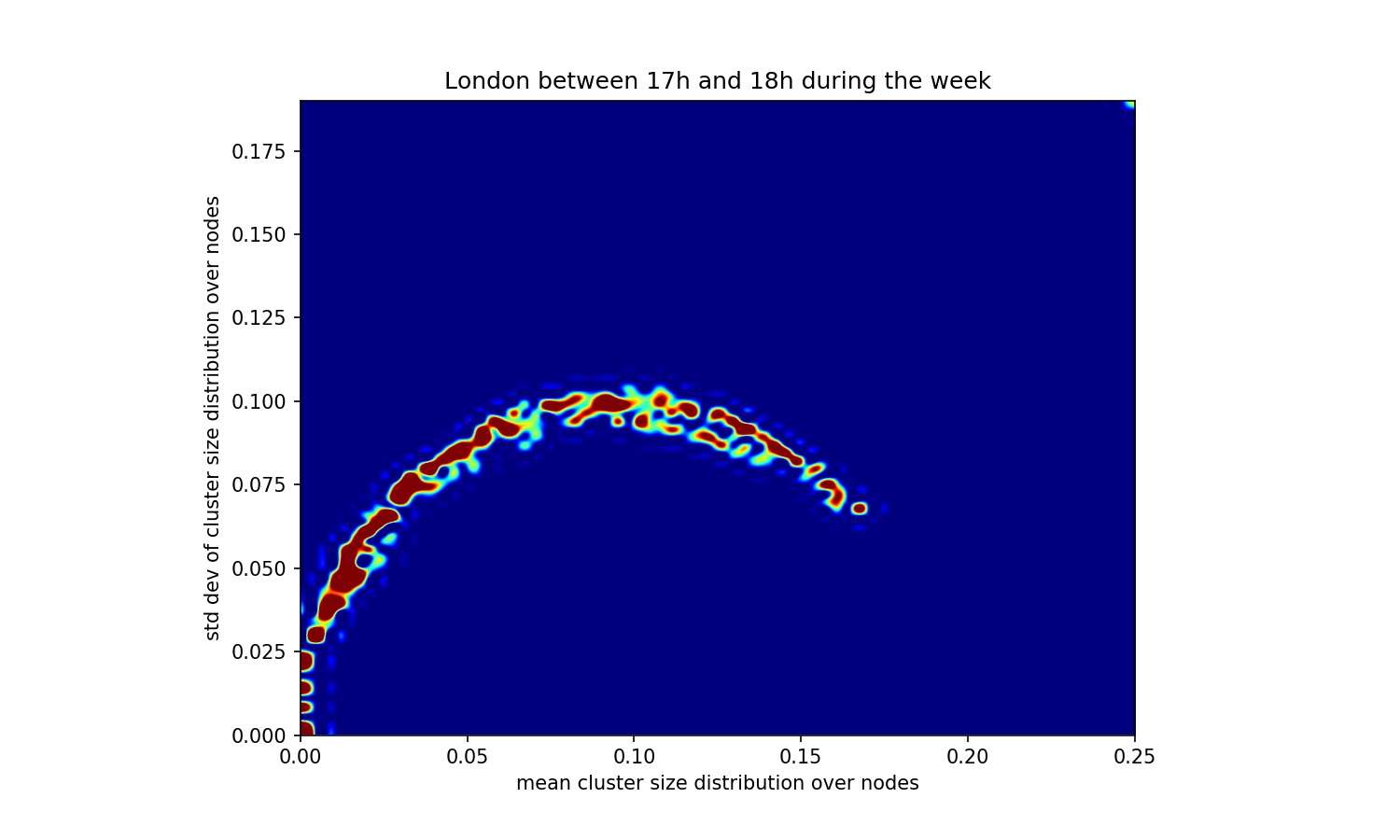}
    \includegraphics[width=0.5\textwidth]{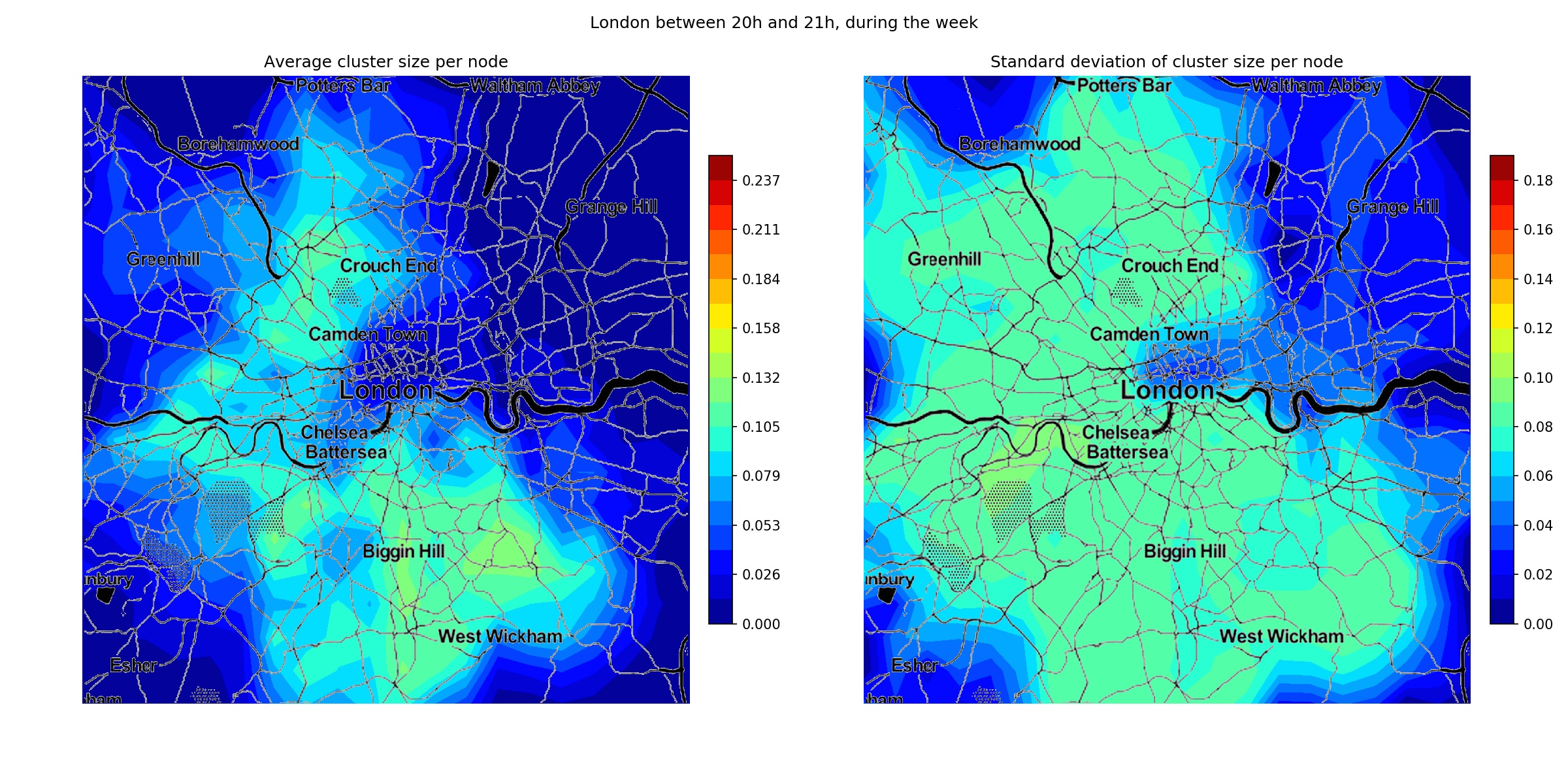}
    \includegraphics[width=0.43\textwidth]{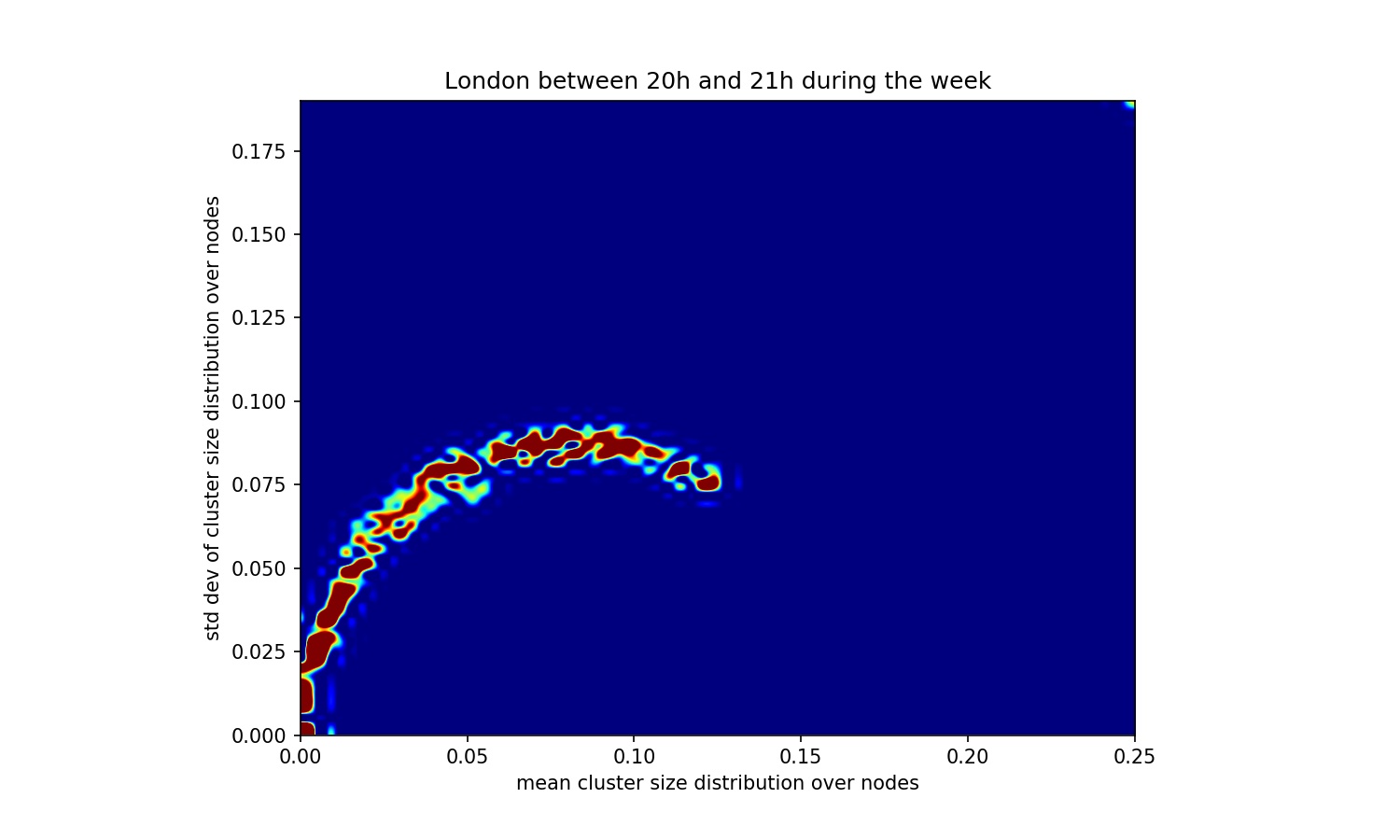}
\caption{Weekdays: Temporally-averaged spatial distributions of the cluster size and standard deviation over London during the first three months of 2019 at four different hours: 8am, 10am, 5pm, and 8pm. On the right, the relation between size and standard deviation over the whole map.}
\label{fig:size-variance-weekdays-london}
\end{figure}

\begin{figure}[H]
\centering
    \includegraphics[width=0.5\textwidth]{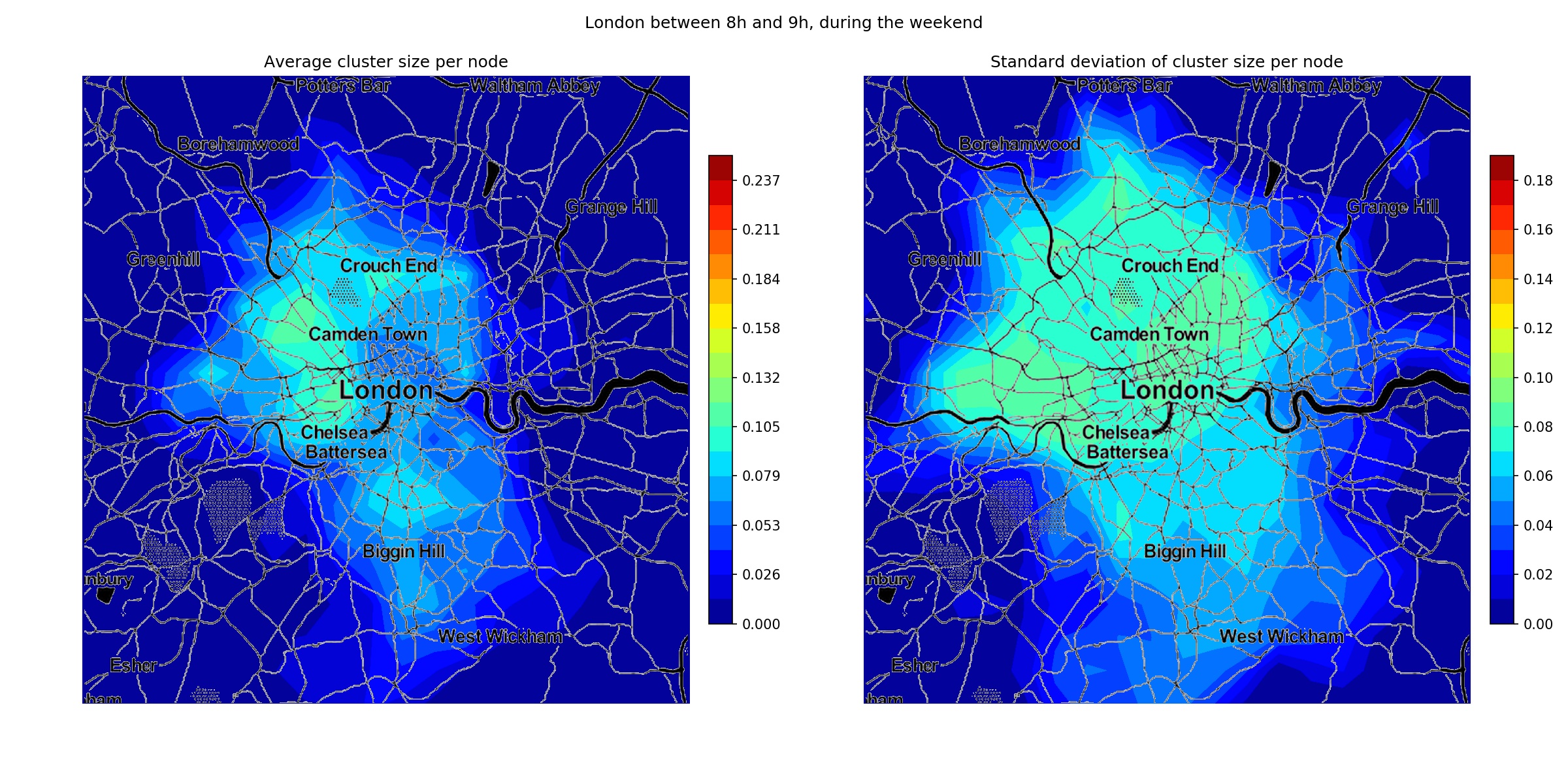}
    \includegraphics[width=0.43\textwidth]{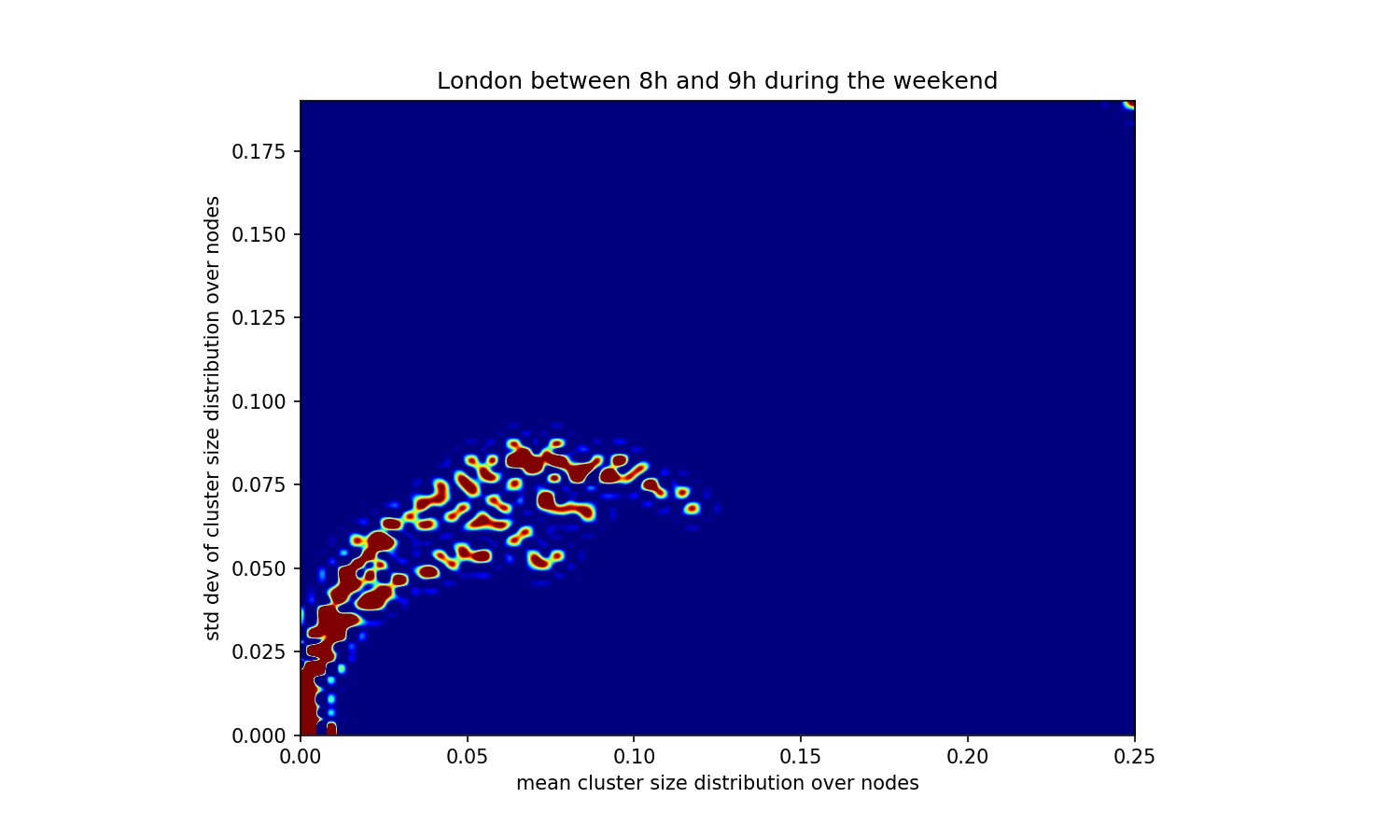}
    \includegraphics[width=0.5\textwidth]{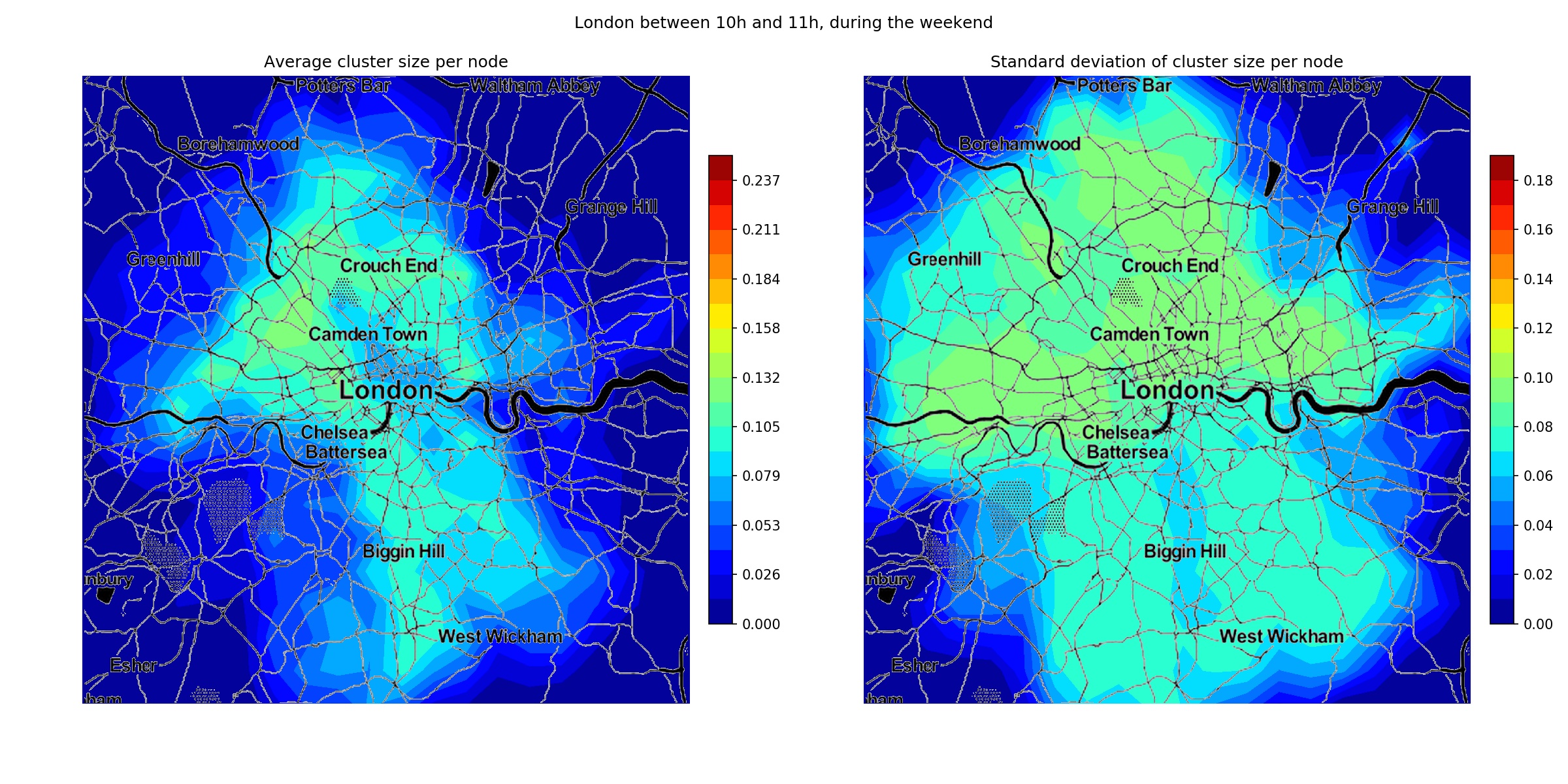}
    \includegraphics[width=0.43\textwidth]{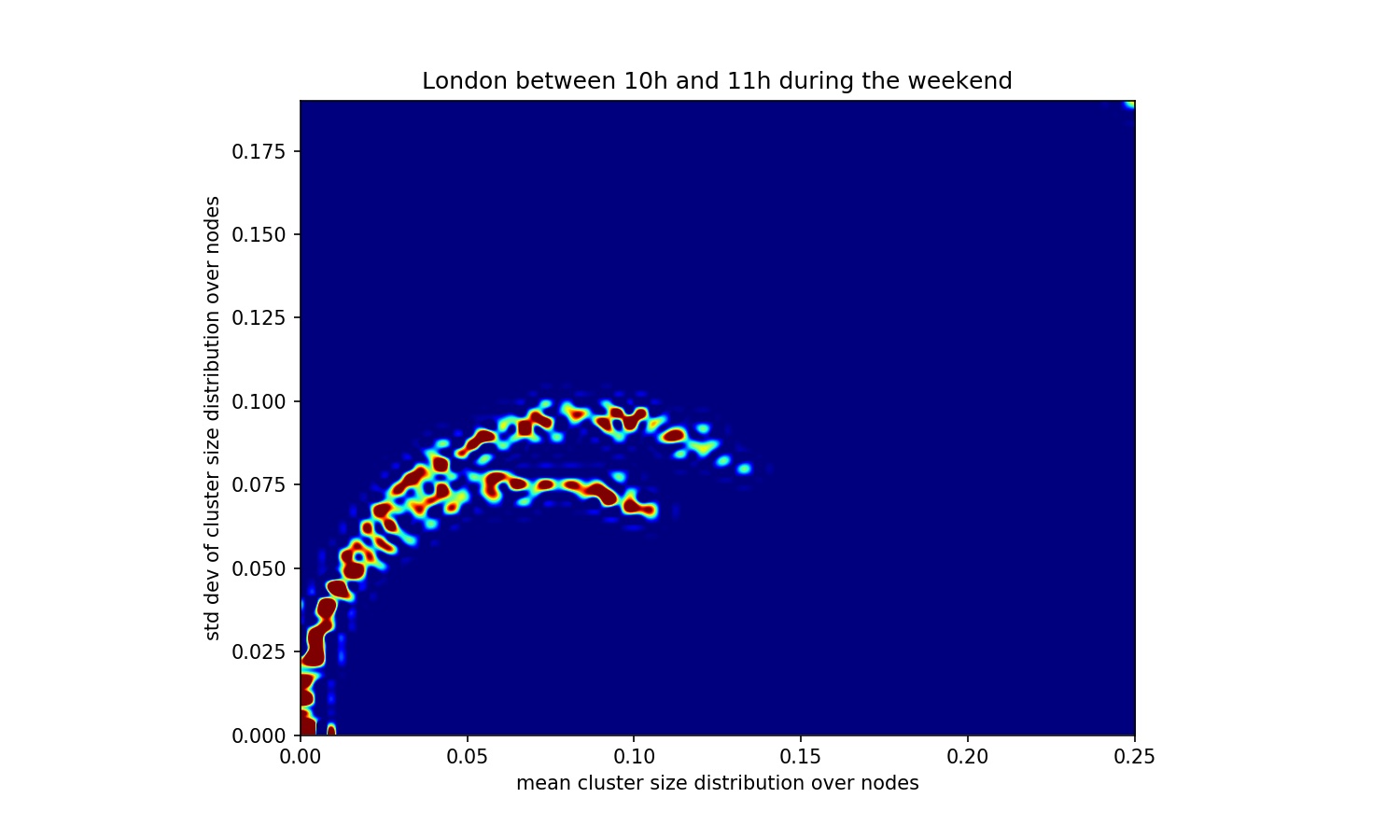}
    \includegraphics[width=0.5\textwidth]{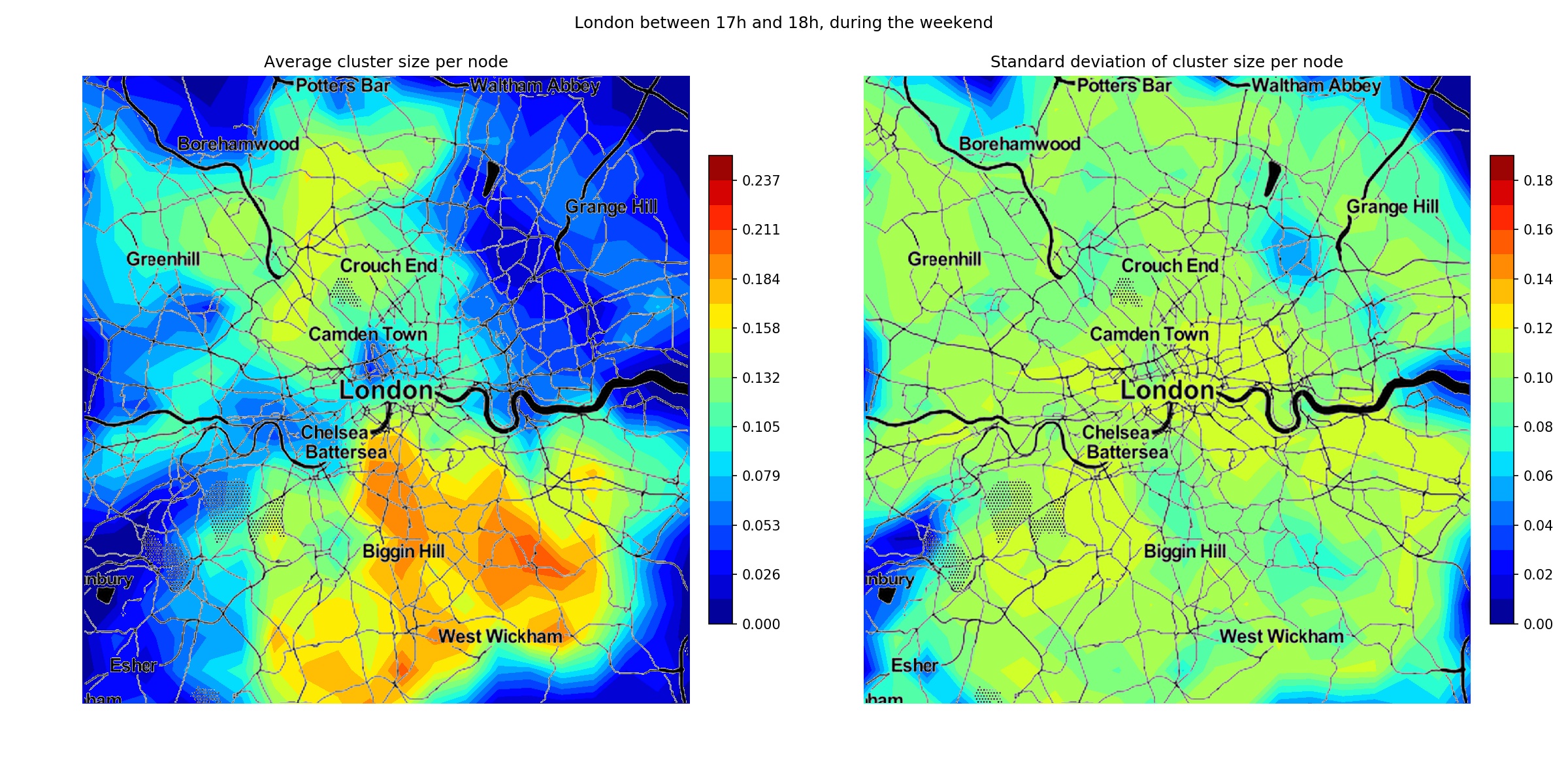}
    \includegraphics[width=0.43\textwidth]{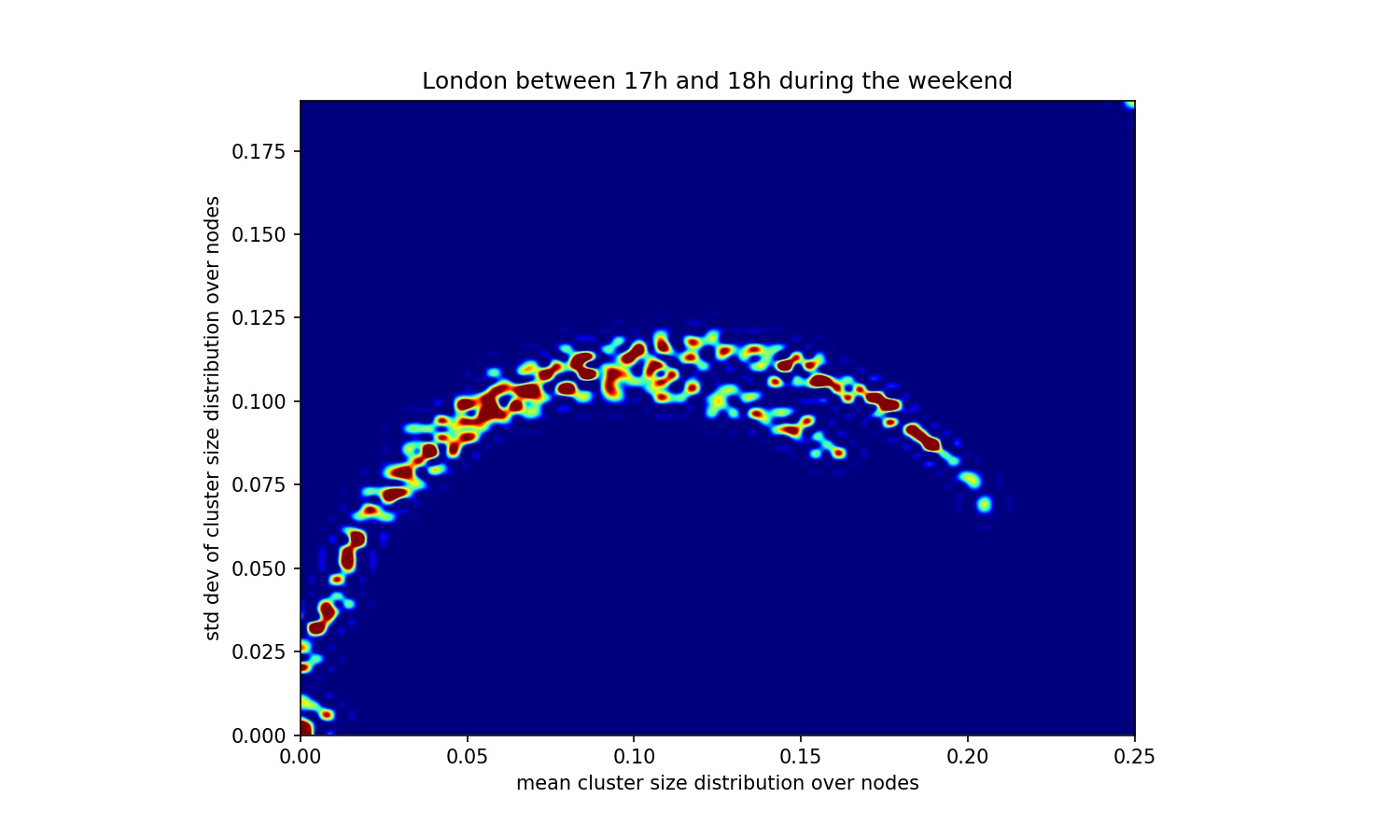}
    \includegraphics[width=0.5\textwidth]{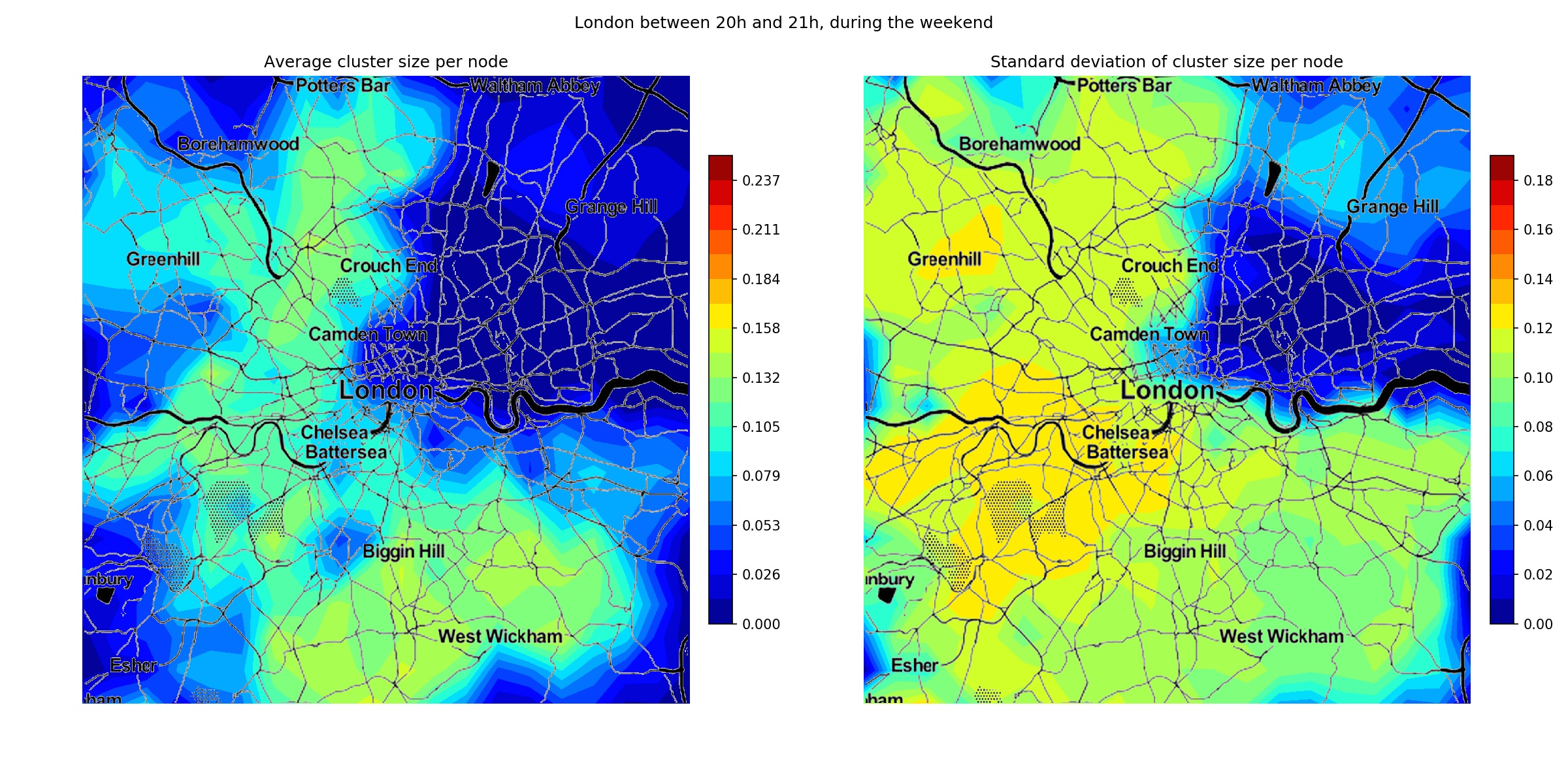}
    \includegraphics[width=0.43\textwidth]{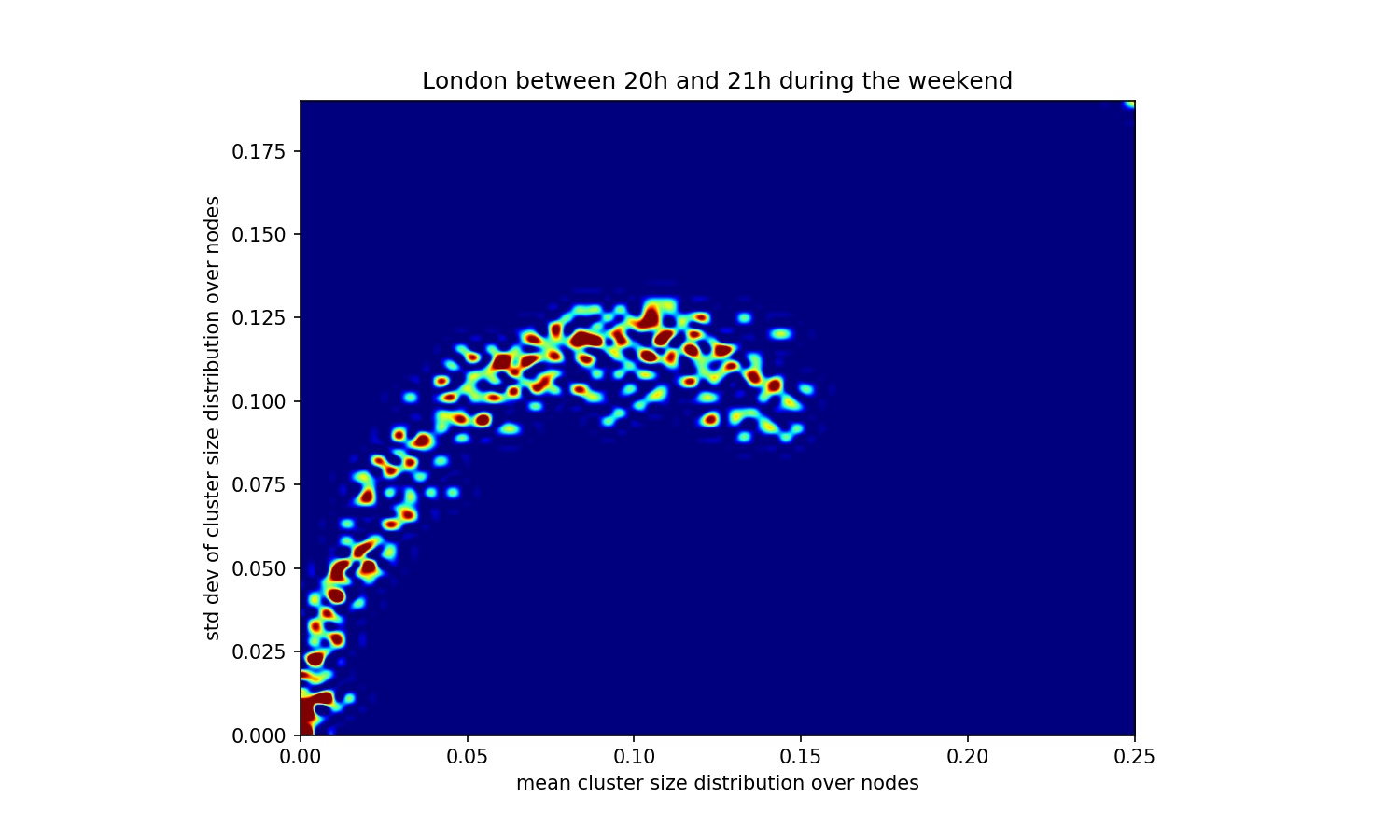}
\caption{Weekends: Temporally-averaged spatial distributions of the cluster size and standard deviation over London during the first three months of 2019 at four different hours: 8am, 10am, 5pm, and 8pm. On the right, the relation between size and standard deviation over the whole map.}
\label{fig:size-variance-weekends-london}
\end{figure}

\clearpage
\section*{Spatial Correlations at Criticality}

\begin{figure}[H]
\centering
    \includegraphics[width=0.7\textwidth]{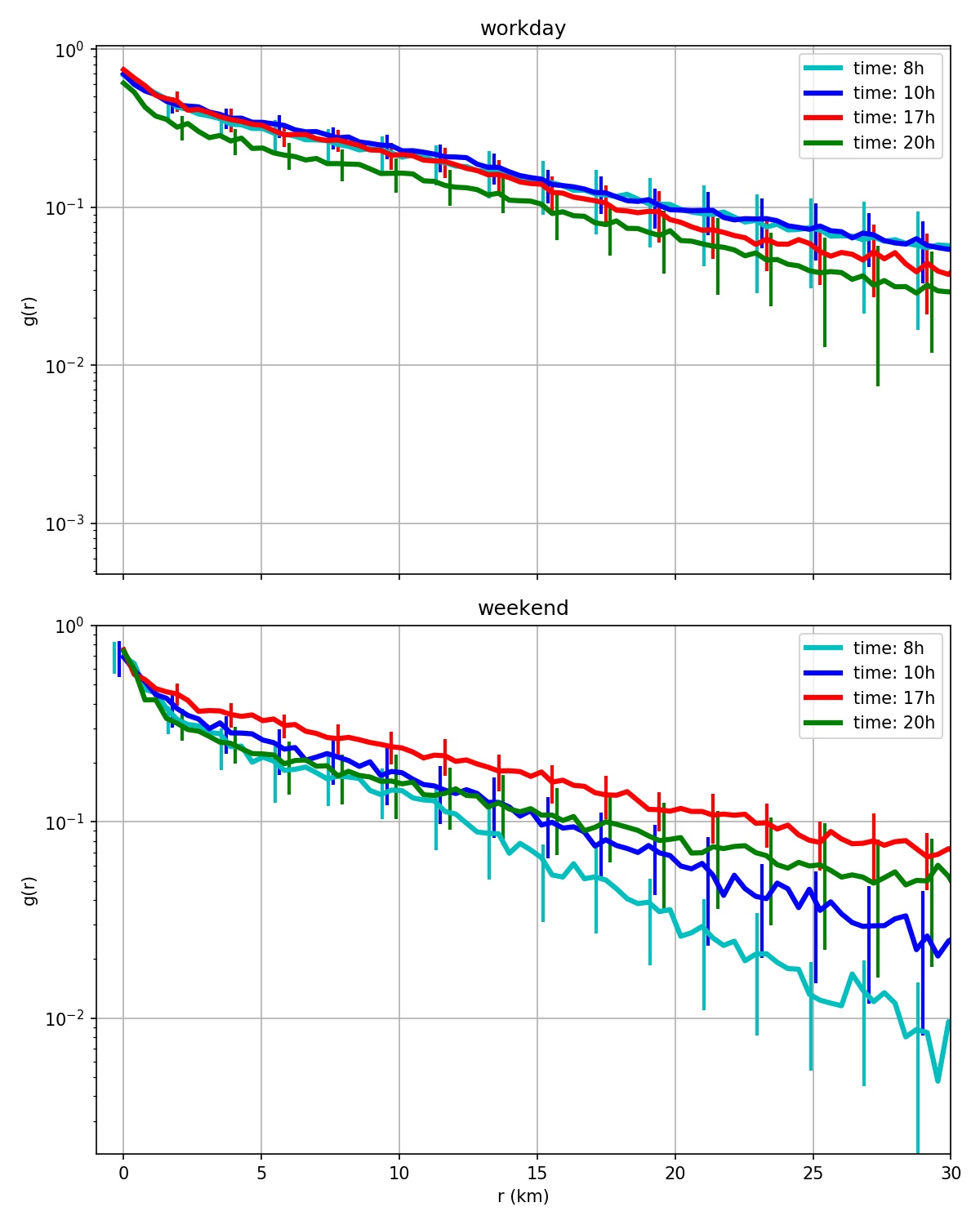}
\caption{Spatial correlations $g(r)$: probability of two edges to belong to the same cluster at distance $r$ for London.}
\label{fig:gdr-london}
\end{figure}

A rough estimate of the behavior of spatial correlation $g(r)$ for London is shown in Fig. \ref{fig:gdr-london}, representing the probability of two edges at distance $r$ of belonging to the same functional cluster, gives an exponential behavior $g(r)\sim e^{-r/\xi}$ with a correlation length $\xi\approx20\pm5$ km of the same order of the considered area.

\clearpage
\renewcommand{\bibnumfmt}[1]{[#1]}
\renewcommand{\citenumfont}[1]{#1}
\section*{UBER Dataset Details} \label{UBER-dataset}
In Refs.~\cite{li_percolation_2015,zeng_switch_2019}, local data interpolation is used to reconstruct missing speed values, but no algorithmic details are given. Here we adopt the following approach: for each edge without speed data, we assign to it the average speed over the edges (with data for the same time interval) that are reachable within a few hops. The number of hops was limited to $5$ to avoid the introduction of non-local artefacts. The median hop length is $70$ m for London and $83$ m for NYC. If no speed data within the selected radius is found, we deem the edge as functional~\cite{li_percolation_2015}, being too far away from intense traffic. The fraction of edges containing a speed value strongly depends on the hour and on the day of the week. Its values range from $0.15$ to $0.5$ for the raw dataset, and from $0.6$ to about $0.9$ after interpolation. To perform the percolation analysis with the UBER dataset, speeds are normalized on each edge, during the relevant hour, by the maximum value observed on that edge over the whole period~\cite{li_percolation_2015}. Normalization is necessary to allow a fair comparison among different road categories, and the resulting normalized speeds will be used to decide whether an edge is functional or not, depending on threshold value.

\begin{figure}[H]
\centering
    \includegraphics[width=0.7\textwidth]{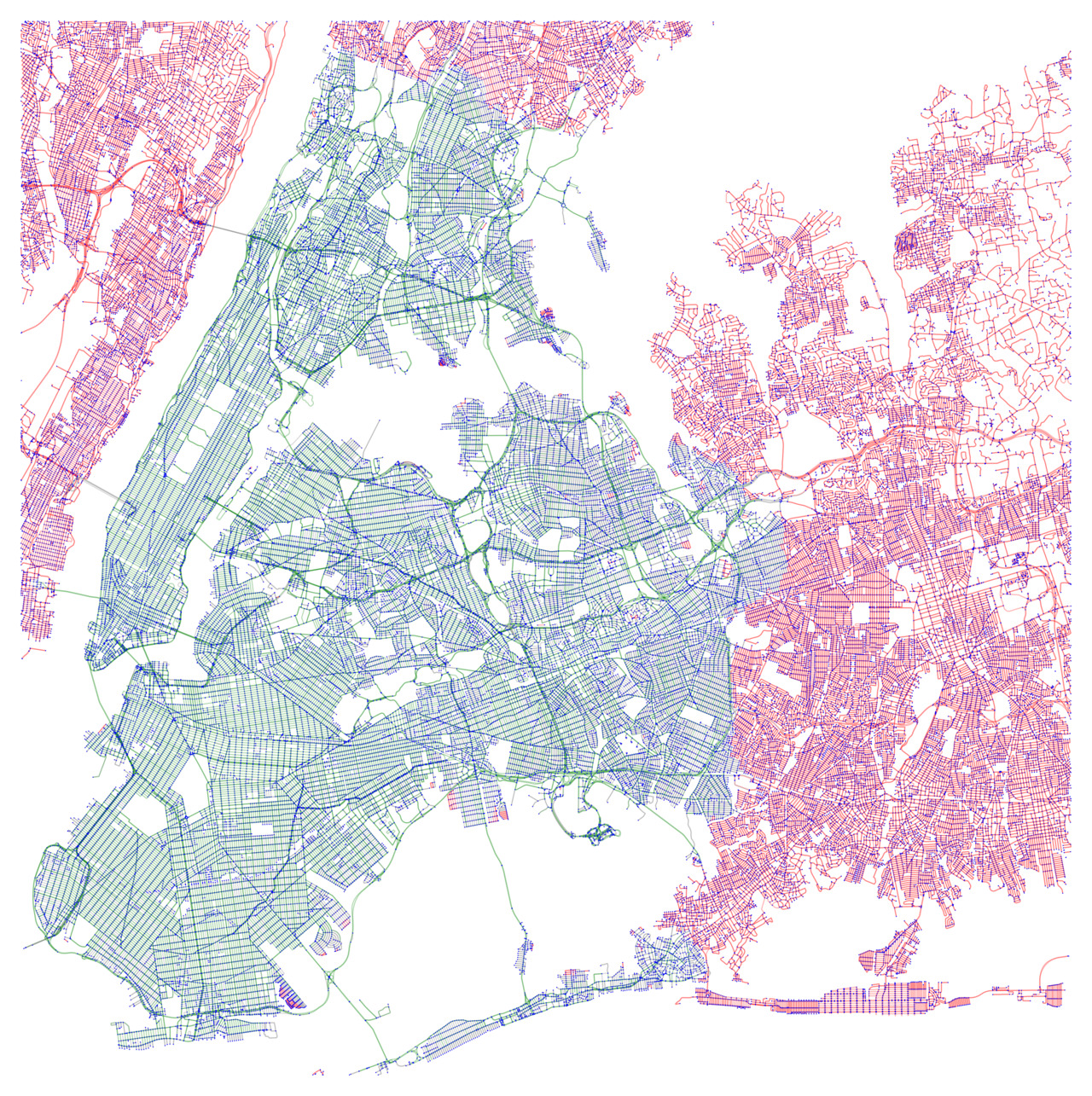}
\caption{NYC map of links present (in green) in the UBER dataset. The red links are never seen in the dataset. State borders are clearly seen. Staten Island, although present in the data was cut off due to poor data quality. Red links were entirely discarded from the graphs analyzed in this work. For comparison, over $95\%$ of the edges present in the London network was found in the UBER dataset at least once.}
\label{fig:map-nyc}
\end{figure}

\end{document}